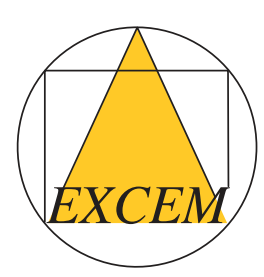

# The Gains, Effective Areas and Equivalent Areas of a Multiport Antenna Array

**FRÉDÉRIC BROYDÉ**[1], **and**
**EVELYNE CLAVELIER**[2]

[1]Eurexcem, 12 chemin des Hauts de Clairefontaine, 78580 Maule, France
[2]Excem, 12 chemin des Hauts de Clairefontaine, 78580 Maule, France

Corresponding author: Frédéric Broydé (e-mail: fredbroyde@eurexcem.com).

**ABSTRACT** We remind the definitions of 22 parameters of a multiport antenna array operating in a specified direction: 4 excitation-dependent parameters for emission, 10 excitation-independent parameters for emission, and 8 parameters for reception. We concisely set forth their main properties. As examples, we compute and discuss these parameters in the cases of two simple 6-port antenna arrays made of parallel dipole antennas. We investigate the effect of a change of variable describing the excitation during emission. This allows us to find an "invariance under a change of excitation variable", and to compare some of the parameters with the ones used by other authors. We study passive linear embeddings of the MAA ports, to reveal an "invariance under an invertible lossless linear embedding".

## I. INTRODUCTION

The selection of real parameters to characterize, specify or optimize a multiport antenna array (MAA) is of considerable practical importance. Indeed, inappropriate parameters will inevitably lead to design and manufacture a suboptimal MAA. This article is about the properties of some recently introduced real MAA parameters.

For emission at a large distance in free space, an MAA may be completely characterized using: a complex matrix such as the scattering matrix or the impedance matrix; and complex vectors often called "embedded element patterns", which depend on a direction in which emission is considered and may be defined in several different ways [1]. In practice, an MAA is often characterized with the absolute values of the entries of its scattering matrix and the absolute values of the entries of a particular choice of embedded element patterns. These real parameters are relevant to the behavior of the MAA used for emission when only one of its ports is excited. However, they do not provide enough information to find out how the MAA radiates in a given direction when several of its ports are excited simultaneously, and they do not allow the behavior of the MAA during reception of a signal incident from this direction to be determined.

To overcome these limitations and better characterize emission and reception of time-harmonic signals by a linear time-invariant (LTI) MAA, [2] defines new parameters that encompass the classic partial gain, absolute gain, effective area and partial effective area for a single-port antenna [3]–[4]. The definitions of the new parameters are based on theoretical results on the unnamed power gains and transducer power gains in multiport circuits, established in [5]–[6].

More precisely, [2] studies emission by the MAA coupled to an arbitrary multiport generator, and reception by the MAA coupled to an arbitrary multiport load. In a specified direction, [2] defines 22 parameters listed in Table 1 to Table 3, which are nonnegative real numbers, the emission parameters being dimensionless gains taking into account simultaneous excitations of all MAA ports, and the reception parameters being effective areas and equivalent areas taking into account the signals delivered by all MAA ports.

In each direction, the 4 gains listed in Table 1 depend on a vector defining the excitation. It is therefore not easy to directly use these gains to compare two MAAs. In contrast, the 10 gains listed in Table 2 are excitation-independent, so that they can be directly used to compare the radiation of two MAAs in a specified direction.

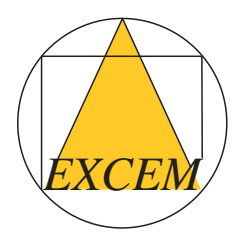

**TABLE 1.** Excitation-dependent parameters for emission.

| Designation | Symbol |
|---|---|
| partial absolute gain | $G_{\mathrm{pa}}(\mathbf{u}_{\mathrm{pol}})$ |
| absolute gain | $G_{\mathrm{a}}$ |
| partial reached gain | $G_{\mathrm{pr}}(\mathbf{u}_{\mathrm{pol}})$ |
| reached gain | $G_{\mathrm{r}}$ |

**TABLE 2.** Excitation-independent parameters for emission.

| Designation | Symbol |
|---|---|
| maximum partial absolute gain | $G_{\mathrm{pa\,MAX}}(\mathbf{u}_{\mathrm{pol}})$ |
| mean partial absolute gain | $G_{\mathrm{pa\,MEA}}(\mathbf{u}_{\mathrm{pol}})$ |
| maximum absolute gain | $G_{\mathrm{a\,MAX}}$ |
| mean absolute gain | $G_{\mathrm{a\,MEA}}$ |
| minimum absolute gain | $G_{\mathrm{a\,MIN}}$ |
| maximum partial reached gain | $G_{\mathrm{pr\,MAX}}(\mathbf{u}_{\mathrm{pol}})$ |
| mean partial reached gain | $G_{\mathrm{pr\,MEA}}(\mathbf{u}_{\mathrm{pol}})$ |
| maximum reached gain | $G_{\mathrm{r\,MAX}}$ |
| mean reached gain | $G_{\mathrm{r\,MEA}}$ |
| minimum reached gain | $G_{\mathrm{r\,MIN}}$ |

**TABLE 3.** Reception parameters.

| Designation | Symbol |
|---|---|
| partial absolute effective area | $A_{\mathrm{pa}}(\mathbf{u}_{\mathrm{pol}})$ |
| absolute effective area | $A_{\mathrm{a}}$ |
| mean absolute equivalent area | $A_{\mathrm{aeq\,MEA}}$ |
| minimum absolute equivalent area | $A_{\mathrm{aeq\,MIN}}$ |
| partial reached effective area | $A_{\mathrm{pr}}(\mathbf{u}_{\mathrm{pol}})$ |
| reached effective area | $A_{\mathrm{r}}$ |
| mean reached equivalent area | $A_{\mathrm{req\,MEA}}$ |
| minimum reached equivalent area | $A_{\mathrm{req\,MIN}}$ |

It is shown in [2] that relationships exist between some of the excitation-independent gains of Table 2 and some of the effective areas and equivalent areas of Table 3, if the MAA is reciprocal. It is also shown in [2] that some parameters of Table 2 and Table 3 can be used in transmission formulas applicable to a radio link between two MAAs. Said relationships and transmission formulas prove the relevance of the new parameters to antenna theory.

This article is meant to show and explain how said 22 parameters of an MAA operating in a specified direction can be used, and to provide some new results. It is organized as follows. Sections II to III are used to remind the definitions of the parameters for emission (gains) and reception (effective areas and equivalent areas). The main properties of the 22 parameters are concisely set forth in Section IV. We compute and discuss these parameters in the cases of two simple 6-port antenna arrays, in Section V. The effect of a change in the variable describing the excitation during emission is examined in Section VI. This allows us to find an "invariance under a change of excitation variable", and to compare some of the parameters with the ones used by other authors. In Section VII, we study passive linear embeddings of the MAA ports, to reveal an "invariance under an invertible lossless linear embedding", which clarifies the possible effects of an ideal decoupling and matching network.

## II. ASSUMPTIONS AND NOTATIONS

Let $\mathbf{M}$ be a complex matrix. $\operatorname{rank}\mathbf{M}$ is the rank of $\mathbf{M}$, $\mathbf{M}^{\mathrm{T}}$ the transpose of $\mathbf{M}$, and $\mathbf{M}^*$ the hermitian adjoint of $\mathbf{M}$. If $\mathbf{M}$ is square, $\operatorname{tr}\mathbf{M}$ is the trace of $\mathbf{M}$ and $H(\mathbf{M})$ is the hermitian part of $\mathbf{M}$ given by $H(\mathbf{M}) = (\mathbf{M} + \mathbf{M}^*)/2$. We use $\overline{\mathbf{M}}$ to denote the complex conjugate of a complex matrix $\mathbf{M}$, so that $\mathbf{M}^* = \overline{\mathbf{M}}^{\mathrm{T}}$. We use $\mathbb{E}$ to denote the Euclidean vector space of dimension 3 associated with physical space. In physical space, we choose a right-handed rectangular cartesian coordinate system $(x, y, z)$ having its origin close to the MAA. A right-handed orthonormal basis $\mathcal{B} = (\mathbf{u}_1, \mathbf{u}_2, \mathbf{u}_3)$ of $\mathbb{E}$ is chosen, and, in what follows, the coordinates of a vector are the coordinates in this basis.

The MAA under study lies in an isotropic, homogeneous and lossless medium. It has $N$ ports numbered from 1 to $N$. It operates in the harmonic steady state, at a wavelength $\lambda$ corresponding to a wave number $k$ and an intrinsic impedance $\eta$ in said medium. The MAA is LTI and passive, but it need not be reciprocal. It has an impedance matrix, denoted by $\mathbf{Z}_{\mathrm{A}}$, of size $N$ by $N$ and such that $H(\mathbf{Z}_{\mathrm{A}})$ is positive definite.

If the MAA is used for emission, an LTI multiport generator having $N$ ports, called MGA, is coupled to the MAA. The ports of the MGA are numbered from 1 to $N$, and, for any integer $p \in \{1, \ldots, N\}$, port $p$ of the MGA is connected to port $p$ of the MAA (positive terminal to positive terminal and negative terminal to negative terminal). We assume that the MGA has an internal impedance matrix $\mathbf{Z}_{\mathrm{G}}$ such that $H(\mathbf{Z}_{\mathrm{G}})$ is positive definite, or equivalently that the MGA has an internal admittance matrix $\mathbf{Y}_{\mathrm{G}}$ such that $H(\mathbf{Y}_{\mathrm{G}})$ is positive definite [7, Sec. IV]. It follows from our assumptions and the explanations provided in [7, Sec. IV–V] that: $\mathbf{Z}_{\mathrm{G}}$ and $\mathbf{Z}_{\mathrm{A}} + \mathbf{Z}_{\mathrm{G}}$ are invertible; and the column vector of the rms currents flowing into ports 1 to $N$ of the MAA, denoted by $\mathbf{I}_{\mathrm{A}}$, can take on any value lying in $\mathbb{C}^N$.

As regards emission, the MAA may be characterized using the matrix $\mathbf{h}_{\mathrm{A}}$ discussed in [2, Sec. III.B], whose columns are the coordinates of the vector effective lengths defined for emission and single-port excitation, the unused ports being left open-circuited. This 3 by $N$ matrix corresponds to a possible set of embedded element patterns [1].

If the MAA is used for reception, an LTI multiport load having $N$ ports, called MLA, is coupled to the MAA. The ports of the MLA are numbered from 1 to $N$, and, for any integer $p \in \{1, \ldots, N\}$, port $p$ of the MLA is connected to port $p$ of the MAA (positive terminal to positive terminal and negative terminal to negative terminal). We assume that the MLA has an impedance matrix $\mathbf{Z}_{\mathrm{L}}$ such that $H(\mathbf{Z}_{\mathrm{L}})$ is positive semidefinite since, if $H(\mathbf{Z}_{\mathrm{L}})$ was not positive semidefinite, the load would not be passive.

As regards reception, the MAA may be characterized using the matrix $\mathbf{h}_{\mathrm{B}}$ discussed in [2, Sec. III.C], whose columns are the coordinates of the vector effective lengths defined for reception using a single port, the unused ports being left open-circuited. $\mathbf{h}_{\mathrm{B}}$ is a 3 by $N$ matrix.

In a given direction, a wave polarization is specified by a polarization vector $\mathbf{u}_{\mathrm{pol}}$, which is a dimensionless complex unit vector explained and discussed in Appendix A.

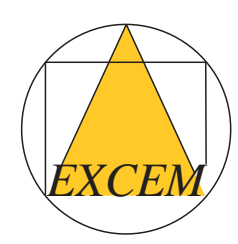

## III. DEFINITIONS

### A. DEFINITIONS ON THE PARTIAL ABSOLUTE GAIN

The partial absolute gain of the MAA, in a given direction, for a given wave polarization and a specified nonzero excitation, is the ratio of a part of the radiation intensity produced by the MAA in the given direction to the radiation intensity that would be obtained if the average power received by the $N$ ports of the MAA were radiated equally in all directions, said part corresponding to the given wave polarization [2, Sec. IV]. For a polarization vector $\mathbf{u}_{\mathrm{pol}}$ and the specified nonzero excitation, the partial absolute gain of the MAA in the given direction is denoted by $G_{\mathrm{pa}}(\mathbf{u}_{\mathrm{pol}})$.

The excitation-independent parameters corresponding to $G_{\mathrm{pa}}(\mathbf{u}_{\mathrm{pol}})$ in the given direction are defined as follows:

- the set of the values of $G_{\mathrm{pa}}(\mathbf{u}_{\mathrm{pol}})$, obtained for all nonzero $\mathbf{I}_{\mathrm{A}}$ has a greatest element referred to as "maximum partial absolute gain" and denoted by $G_{\mathrm{pa\,MAX}}(\mathbf{u}_{\mathrm{pol}})$;
- the quantity $G_{\mathrm{pa\,MEA}}(\mathbf{u}_{\mathrm{pol}}) = G_{\mathrm{pa\,MAX}}(\mathbf{u}_{\mathrm{pol}})/N$ is referred to as "mean partial absolute gain" because it is a mean value of $G_{\mathrm{pa}}(\mathbf{u}_{\mathrm{pol}})$ over a number $N$ of linearly independent $\mathbf{I}_{\mathrm{A}}$.

### B. DEFINITIONS ON THE ABSOLUTE GAIN

The absolute gain of the MAA, in a given direction, for a specified nonzero excitation, is the ratio of the radiation intensity produced by the MAA in the given direction to the radiation intensity that would be obtained if the average power received by the $N$ ports of the MAA were radiated equally in all directions [2, Sec. V]. The absolute gain of the MAA in the given direction for the specified nonzero excitation is denoted by $G_{\mathrm{a}}$.

The excitation-independent parameters corresponding to $G_{\mathrm{a}}$ in the given direction are defined as follows:

- the set of the values of $G_{\mathrm{a}}$, obtained for all nonzero $\mathbf{I}_{\mathrm{A}}$ has a least element referred to as "minimum absolute gain" and denoted by $G_{\mathrm{a\,MIN}}$, and a greatest element referred to as "maximum absolute gain" and denoted by $G_{\mathrm{a\,MAX}}$;
- the $N$ by $N$ matrix
$$\mathbf{N}_{\mathrm{A}} = \frac{\pi\eta}{\lambda^2}\,\mathbf{h}_{\mathrm{A}}^{*}\mathbf{h}_{\mathrm{A}} \tag{1}$$
is positive semidefinite, $\mathrm{rank}\,\mathbf{N}_{\mathrm{A}} \leqslant 2$ and the "mean absolute gain" denoted by $G_{\mathrm{a\,MEA}}$ is defined as
$$G_{\mathrm{a\,MEA}} = \frac{\mathrm{tr}\left(\mathbf{N}_{\mathrm{A}} H(\mathbf{Z}_{\mathrm{A}})^{-1}\right)}{N}\,, \tag{2}$$
because this positive real quantity is a mean value of $G_{\mathrm{a}}$ over a number $N$ of linearly independent $\mathbf{I}_{\mathrm{A}}$.

### C. DEFINITIONS ON THE PARTIAL REACHED GAIN

The partial reached gain of the MAA, in a given direction, for a given wave polarization and a specified nonzero excitation, is the ratio of a part of the radiation intensity produced by the MAA in the given direction to the radiation intensity that would be obtained if the available power of the MGA coupled to the $N$ ports of the MAA were radiated equally in all directions, said part corresponding to the given wave polarization [2, Sec. VII]. For a polarization vector $\mathbf{u}_{\mathrm{pol}}$ and the specified nonzero excitation, the partial reached gain of the MAA in the given direction is denoted by $G_{\mathrm{pr}}(\mathbf{u}_{\mathrm{pol}})$.

The excitation-independent parameters corresponding to $G_{\mathrm{pr}}(\mathbf{u}_{\mathrm{pol}})$ in the given direction are defined as follows:

- the set of the values of $G_{\mathrm{pr}}(\mathbf{u}_{\mathrm{pol}})$, obtained for all nonzero $\mathbf{I}_{\mathrm{A}}$ has a greatest element referred to as "maximum partial reached gain" and denoted by $G_{\mathrm{pr\,MAX}}(\mathbf{u}_{\mathrm{pol}})$;
- the quantity $G_{\mathrm{pr\,MEA}}(\mathbf{u}_{\mathrm{pol}}) = G_{\mathrm{pr\,MAX}}(\mathbf{u}_{\mathrm{pol}})/N$ is referred to as "mean partial reached gain" because it is a mean value of $G_{\mathrm{pr}}(\mathbf{u}_{\mathrm{pol}})$ over a number $N$ of linearly independent $\mathbf{I}_{\mathrm{A}}$.

### D. DEFINITIONS ON THE REACHED GAIN

The reached gain of the MAA, in a given direction, for a specified nonzero excitation, is the ratio of the radiation intensity produced by the MAA in the given direction to the radiation intensity that would be obtained if the available power of the MGA coupled to the $N$ ports of the MAA were radiated equally in all directions [2, Sec. VIII]. For the specified nonzero excitation, the reached gain of the MAA in the given direction is denoted by $G_{\mathrm{r}}$.

The excitation-independent parameters corresponding to $G_{\mathrm{r}}$ in the given direction are defined as follows:

- the set of the values of $G_{\mathrm{r}}$, obtained for all nonzero $\mathbf{I}_{\mathrm{A}}$ has a least element referred to as "minimum reached gain" and denoted by $G_{\mathrm{r\,MIN}}$, and a greatest element referred to as "maximum reached gain" and denoted by $G_{\mathrm{r\,MAX}}$;
- the $N$ by $N$ impedance matrix
$$\mathbf{Z}_{\mathrm{AVG}} = (\mathbf{Z}_{\mathrm{A}} + \mathbf{Z}_{\mathrm{G}})^{*}\,\frac{H(\mathbf{Z}_{\mathrm{G}})^{-1}}{4}\,(\mathbf{Z}_{\mathrm{A}} + \mathbf{Z}_{\mathrm{G}}) \tag{3}$$
is positive definite, and the "mean reached gain" denoted by $G_{\mathrm{r\,MEA}}$ is defined as
$$G_{\mathrm{r\,MEA}} = \frac{\mathrm{tr}\left(\mathbf{N}_{\mathrm{A}} \mathbf{Z}_{\mathrm{AVG}}^{-1}\right)}{N}\,, \tag{4}$$
because this positive real quantity is a mean value of $G_{\mathrm{r}}$ over a number $N$ of linearly independent $\mathbf{I}_{\mathrm{A}}$.

### E. ABSOLUTE RECEPTION PARAMETERS

The partial absolute effective area of the MAA, in a given direction, for a given wave polarization, is the ratio of the available power at the $N$ ports of the MAA used for reception to the power flux density of a uniform plane wave of the given polarization incident on the MAA from the given direction [2, Sec. IV]. The partial absolute effective area of the MAA in the given direction for a polarization vector $\mathbf{u}_{\mathrm{pol}}$ is denoted by $A_{\mathrm{pa}}(\mathbf{u}_{\mathrm{pol}})$.

The absolute effective area of the MAA, in a given direction, is the ratio of the available power at the $N$ ports of the MAA used for reception to the power flux density of a uniform plane wave incident on the MAA from the given direction, the polarization of the uniform plane wave being such that it maximizes this available power [2, Sec. V]. The absolute effective area of the MAA in the given direction is denoted by $A_{\mathrm{a}}$. In a given direction, $A_{\mathrm{a}}$ is equal to $A_{\mathrm{pa}}(\mathbf{u}_{\mathrm{pol}})$ for a special value of $\mathbf{u}_{\mathrm{pol}}$.

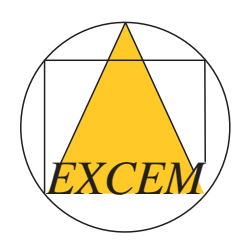

The minimum absolute equivalent area of the MAA, in a given direction, is the ratio of the available power at the $N$ ports of the MAA used for reception to the power flux density of a uniform plane wave incident on the MAA from the given direction, the polarization of the uniform plane wave being such that it minimizes this available power [2, Sec. VI]. The minimum absolute equivalent area of the MAA in the given direction is denoted by $A_{\mathrm{aeq\,MIN}}$. In a given direction, $A_{\mathrm{aeq\,MIN}}$ is equal to $A_{\mathrm{pa}}(\mathbf{u}_{\mathrm{pol}})$ for a special value of $\mathbf{u}_{\mathrm{pol}}$.

In a given direction, $A_{\mathrm{aeq\,MEA}} = (A_{\mathrm{a}} + A_{\mathrm{aeq\,MIN}})/2$ is referred to as "mean absolute equivalent area" because it is a mean value of $A_{\mathrm{pa}}(\mathbf{u}_{\mathrm{pol}})$ over two orthogonal polarization vectors.

### F. REACHED RECEPTION PARAMETERS

The partial reached effective area of the MAA, in a given direction, for a given wave polarization, is the ratio of the average power delivered by the $N$ ports of the MAA used for reception to the power flux density of a uniform plane wave of the given polarization incident on the MAA from the given direction [2, Sec. VII]. The partial reached effective area of the MAA in the given direction for a polarization vector $\mathbf{u}_{\mathrm{pol}}$ is denoted by $A_{\mathrm{pr}}(\mathbf{u}_{\mathrm{pol}})$.

The reached effective area of the MAA, in a given direction, is the ratio of the average power delivered by the $N$ ports of the MAA used for reception to the power flux density of a uniform plane wave incident on the MAA from the given direction, the polarization of the uniform plane wave being such that it maximizes this average power [2, Sec. VIII]. The reached effective area of the MAA in the given direction is denoted by $A_{\mathrm{r}}$. In a given direction, $A_{\mathrm{r}}$ is equal to $A_{\mathrm{pr}}(\mathbf{u}_{\mathrm{pol}})$ for a special value of $\mathbf{u}_{\mathrm{pol}}$.

The minimum reached equivalent area of the MAA, in a given direction, is the ratio of the average power delivered by the $N$ ports of the MAA used for reception to the power flux density of a uniform plane wave incident on the MAA from the given direction, the polarization of the uniform plane wave being such that it minimizes this average power [2, Sec. IX]. The minimum reached equivalent area of the MAA in the given direction is denoted by $A_{\mathrm{req\,MIN}}$. In a given direction, $A_{\mathrm{req\,MIN}}$ is equal to $A_{\mathrm{pr}}(\mathbf{u}_{\mathrm{pol}})$ for a special value of $\mathbf{u}_{\mathrm{pol}}$.

In a given direction, $A_{\mathrm{req\,MEA}} = (A_{\mathrm{r}} + A_{\mathrm{req\,MIN}})/2$ is referred to as "mean reached equivalent area" because it is a mean value of $A_{\mathrm{pr}}(\mathbf{u}_{\mathrm{pol}})$ over two orthogonal polarization vectors.

## IV. MAIN PROPERTIES

### A. ORIGIN OF THE STATED RESULTS

The results stated without explanation in this Section IV are directly based on [2, Sec. IV-IX and XI], and also on [8] for the results involving a "surface-element-ratio probability density".

In [2, Sec. III.C], we assumed that $\mathbf{Z}_{\mathrm{G}} = \mathbf{Z}_{\mathrm{L}}$ to avoid repeating this assumption where it was needed. Above in Section II, this assumption was not used, so that (13)–(14) are different from their counterparts in [2].

### B. PROPERTIES RELATING TO THE GAINS

We use $\underset{\sim}{\mathbf{u}}_{\mathrm{pol}}$ to denote the coordinates of $\mathbf{u}_{\mathrm{pol}}$ in the chosen orthonormal basis $\mathcal{B}$ of $\mathbb{E}$, and we define the $N$ by $N$ matrix

$$\mathbf{N}_{\mathrm{Ap}}(\mathbf{u}_{\mathrm{pol}}) = \frac{\pi\eta}{\lambda^2}\, \mathbf{h}_{\mathrm{A}}^{*}\, \overline{\underset{\sim}{\mathbf{u}}_{\mathrm{pol}}}\, \underset{\sim}{\mathbf{u}}_{\mathrm{pol}}^{\mathrm{T}}\, \mathbf{h}_{\mathrm{A}}\,, \tag{5}$$

which is positive semidefinite.

$G_{\mathrm{pa}}(\mathbf{u}_{\mathrm{pol}})$ is a generalized Rayleigh ratio of $\mathbf{N}_{\mathrm{Ap}}(\mathbf{u}_{\mathrm{pol}})$ to $H(\mathbf{Z}_{\mathrm{A}})$, in the variable $\mathbf{I}_{\mathrm{A}}$, that is to say: if $\mathbf{I}_{\mathrm{A}} \neq \mathbf{0}$, then

$$G_{\mathrm{pa}}(\mathbf{u}_{\mathrm{pol}}) = \frac{\mathbf{I}_{\mathrm{A}}^{*}\, \mathbf{N}_{\mathrm{Ap}}(\mathbf{u}_{\mathrm{pol}})\, \mathbf{I}_{\mathrm{A}}}{\mathbf{I}_{\mathrm{A}}^{*}\, H(\mathbf{Z}_{\mathrm{A}})\, \mathbf{I}_{\mathrm{A}}}\,. \tag{6}$$

The eigenvalues of $\mathbf{N}_{\mathrm{Ap}}(\mathbf{u}_{\mathrm{pol}})\, H(\mathbf{Z}_{\mathrm{A}})^{-1}$ are real and nonnegative, and $G_{\mathrm{pa\,MAX}}(\mathbf{u}_{\mathrm{pol}})$ is the largest of them.

We have $G_{\mathrm{pa\,MEA}}(\mathbf{u}_{\mathrm{pol}}) = \mathrm{tr}\,(\mathbf{N}_{\mathrm{Ap}}(\mathbf{u}_{\mathrm{pol}})\, H(\mathbf{Z}_{\mathrm{A}})^{-1})/N$ because $\mathrm{rank}\, \mathbf{N}_{\mathrm{Ap}}(\mathbf{u}_{\mathrm{pol}}) \leqslant 1$. Thus, in a given direction, $G_{\mathrm{pa\,MEA}}(\mathbf{u}_{\mathrm{pol}})$ is the expected value of $G_{\mathrm{pa}}(\mathbf{u}_{\mathrm{pol}})$ if $\mathbf{I}_{\mathrm{A}}$ is random and distributed according to a surface-element-ratio probability density that only depends on $H(\mathbf{Z}_{\mathrm{A}})$.

$G_{\mathrm{a}}$ is a generalized Rayleigh ratio of $\mathbf{N}_{\mathrm{A}}$ to $H(\mathbf{Z}_{\mathrm{A}})$, in the variable $\mathbf{I}_{\mathrm{A}}$, that is to say: if $\mathbf{I}_{\mathrm{A}} \neq \mathbf{0}$, then

$$G_{\mathrm{a}} = \frac{\mathbf{I}_{\mathrm{A}}^{*} \mathbf{N}_{\mathrm{A}}\, \mathbf{I}_{\mathrm{A}}}{\mathbf{I}_{\mathrm{A}}^{*}\, H(\mathbf{Z}_{\mathrm{A}})\, \mathbf{I}_{\mathrm{A}}}\,. \tag{7}$$

The eigenvalues of $\mathbf{N}_{\mathrm{A}} H(\mathbf{Z}_{\mathrm{A}})^{-1}$ are real and nonnegative, $G_{\mathrm{a\,MAX}}$ is the largest of these eigenvalues, and $G_{\mathrm{a\,MIN}}$ is the least of these eigenvalues. It follows from (2) that, in a given direction, $G_{\mathrm{a\,MEA}}$ is the expected value of $G_{\mathrm{a}}$ if $\mathbf{I}_{\mathrm{A}}$ is random and distributed according to said surface-element-ratio probability density that only depends on $H(\mathbf{Z}_{\mathrm{A}})$.

$G_{\mathrm{pr}}(\mathbf{u}_{\mathrm{pol}})$ is a generalized Rayleigh ratio of $\mathbf{N}_{\mathrm{Ap}}(\mathbf{u}_{\mathrm{pol}})$ to $\mathbf{Z}_{\mathrm{AVG}}$, in the variable $\mathbf{I}_{\mathrm{A}}$, that is to say: if $\mathbf{I}_{\mathrm{A}} \neq \mathbf{0}$, then

$$G_{\mathrm{pr}}(\mathbf{u}_{\mathrm{pol}}) = \frac{\mathbf{I}_{\mathrm{A}}^{*}\, \mathbf{N}_{\mathrm{Ap}}(\mathbf{u}_{\mathrm{pol}})\, \mathbf{I}_{\mathrm{A}}}{\mathbf{I}_{\mathrm{A}}^{*}\, \mathbf{Z}_{\mathrm{AVG}}\, \mathbf{I}_{\mathrm{A}}}\,. \tag{8}$$

The eigenvalues of $\mathbf{N}_{\mathrm{Ap}}(\mathbf{u}_{\mathrm{pol}})\, \mathbf{Z}_{\mathrm{AVG}}^{-1}$ are real and nonnegative, and $G_{\mathrm{pr\,MAX}}(\mathbf{u}_{\mathrm{pol}})$ is the largest of them.

We have $G_{\mathrm{pr\,MEA}}(\mathbf{u}_{\mathrm{pol}}) = \mathrm{tr}\,(\mathbf{N}_{\mathrm{Ap}}(\mathbf{u}_{\mathrm{pol}})\, \mathbf{Z}_{\mathrm{AVG}}^{-1})/N$ because $\mathrm{rank}\, \mathbf{N}_{\mathrm{Ap}}(\mathbf{u}_{\mathrm{pol}}) \leqslant 1$. Thus, in a given direction, $G_{\mathrm{pr\,MEA}}(\mathbf{u}_{\mathrm{pol}})$ is the expected value of $G_{\mathrm{pr}}(\mathbf{u}_{\mathrm{pol}})$ if $\mathbf{I}_{\mathrm{A}}$ is random and distributed according to a surface-element-ratio probability density that only depends on $\mathbf{Z}_{\mathrm{AVG}}$.

$G_{\mathrm{r}}$ is a generalized Rayleigh ratio of $\mathbf{N}_{\mathrm{A}}$ to $\mathbf{Z}_{\mathrm{AVG}}$, in the variable $\mathbf{I}_{\mathrm{A}}$, that is to say: if $\mathbf{I}_{\mathrm{A}} \neq \mathbf{0}$, then

$$G_{\mathrm{r}} = \frac{\mathbf{I}_{\mathrm{A}}^{*} \mathbf{N}_{\mathrm{A}}\, \mathbf{I}_{\mathrm{A}}}{\mathbf{I}_{\mathrm{A}}^{*}\, \mathbf{Z}_{\mathrm{AVG}}\, \mathbf{I}_{\mathrm{A}}}\,. \tag{9}$$

The eigenvalues of $\mathbf{N}_{\mathrm{A}} \mathbf{Z}_{\mathrm{AVG}}^{-1}$ are real and nonnegative, $G_{\mathrm{r\,MAX}}$ is the largest of these eigenvalues, and $G_{\mathrm{r\,MIN}}$ is the least of these eigenvalues. It follows from (4) that, in a given direction, $G_{\mathrm{r\,MEA}}$ is the expected value of $G_{\mathrm{r}}$ if $\mathbf{I}_{\mathrm{A}}$ is random and distributed according to said surface-element-ratio probability density that only depends on $\mathbf{Z}_{\mathrm{AVG}}$.

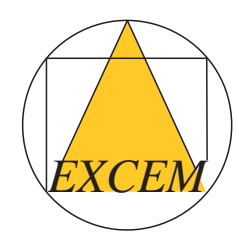

### C. ON THE EFFECTIVE AND EQUIVALENT AREAS

To compute $A_{\mathrm{pa}}(\mathbf{u}_{\mathrm{pol}})$, we can use

$$A_{\mathrm{pa}}(\mathbf{u}_{\mathrm{pol}}) = \frac{\eta}{4}\, \mathbf{u}^*_{\mathrm{pol}} \overline{\mathbf{h}_{\mathrm{B}}}\, H(\mathbf{Z}_{\mathrm{A}})^{-1} \mathbf{h}^{\mathrm{T}}_{\mathrm{B}}\, \mathbf{u}_{\mathrm{pol}} \,. \tag{10}$$

Let

$$\mathbf{P} = \begin{pmatrix} 0 & 0 \\ 1 & 0 \\ 0 & 1 \end{pmatrix} . \tag{11}$$

If the chosen basis $\mathcal{B} = (\mathbf{u}_1, \mathbf{u}_2, \mathbf{u}_3)$ of $\mathbb{E}$ is such that $\mathbf{u}_1$ is pointing in the direction from which a uniform plane wave is incident on the MAA, we use $\mathbf{N}_{\mathrm{Ba}}$ to denote the 2 by 2 matrix given by

$$\mathbf{N}_{\mathrm{Ba}} = \frac{\eta}{4}\, \mathbf{P}^{\mathrm{T}} \overline{\mathbf{h}_{\mathrm{B}}}\, H(\mathbf{Z}_{\mathrm{A}})^{-1} \mathbf{h}^{\mathrm{T}}_{\mathrm{B}}\, \mathbf{P} \,, \tag{12}$$

which is positive semidefinite. The eigenvalues of $\mathbf{N}_{\mathrm{Ba}}$ are real and nonnegative, $A_{\mathrm{a}}$ is the largest of these eigenvalues, and $A_{\mathrm{aeq\,MIN}}$ is the least of these eigenvalues. We have $A_{\mathrm{aeq\,MEA}} = \operatorname{tr} \mathbf{N}_{\mathrm{Ba}}/2$. Moreover, in a given direction, $A_{\mathrm{aeq\,MEA}}$ is the expected value of $A_{\mathrm{pa}}(\mathbf{u}_{\mathrm{pol}})$ for a random polarization having a uniform probability density.

To compute $A_{\mathrm{pr}}(\mathbf{u}_{\mathrm{pol}})$, we can use

$$A_{\mathrm{pr}}(\mathbf{u}_{\mathrm{pol}}) = \eta\, \mathbf{u}^*_{\mathrm{pol}} \overline{\mathbf{h}_{\mathrm{B}}}\, \mathbf{Y}^*_{\mathrm{AL}} H(\mathbf{Z}_{\mathrm{L}}) \mathbf{Y}_{\mathrm{AL}} \mathbf{h}^{\mathrm{T}}_{\mathrm{B}}\, \mathbf{u}_{\mathrm{pol}} \,, \tag{13}$$

where $\mathbf{Y}_{\mathrm{AL}} = (\mathbf{Z}_{\mathrm{A}} + \mathbf{Z}_{\mathrm{L}})^{-1}$.

If the chosen basis $\mathcal{B} = (\mathbf{u}_1, \mathbf{u}_2, \mathbf{u}_3)$ is such that $\mathbf{u}_1$ is pointing in the direction from which a uniform plane wave is incident on the MAA, we use $\mathbf{N}_{\mathrm{Br}}$ to denote the 2 by 2 matrix given by

$$\mathbf{N}_{\mathrm{Br}} = \eta\, \mathbf{P}^{\mathrm{T}} \overline{\mathbf{h}_{\mathrm{B}}}\, \mathbf{Y}^*_{\mathrm{AL}} H(\mathbf{Z}_{\mathrm{L}}) \mathbf{Y}_{\mathrm{AL}} \mathbf{h}^{\mathrm{T}}_{\mathrm{B}}\, \mathbf{P} \,, \tag{14}$$

which is positive semidefinite. The eigenvalues of $\mathbf{N}_{\mathrm{Br}}$ are real and nonnegative, $A_{\mathrm{r}}$ is the largest of these eigenvalues, and $A_{\mathrm{req\,MIN}}$ is the least of these eigenvalues. We have $A_{\mathrm{req\,MEA}} = \operatorname{tr} \mathbf{N}_{\mathrm{Br}}/2$. Moreover, in a given direction, $A_{\mathrm{req\,MEA}}$ is the expected value of $A_{\mathrm{pr}}(\mathbf{u}_{\mathrm{pol}})$ for a random polarization having a uniform probability density.

### D. RELATIONSHIPS FOR A RECIPROCAL MAA

If the MAA is reciprocal, then $\mathbf{Z}_{\mathrm{A}}$ is symmetric, $\mathbf{h}_{\mathrm{B}} = \mathbf{h}_{\mathrm{A}}$ and we can assert that, in a given direction:

$$A_{\mathrm{pa}}(\mathbf{u}_{\mathrm{pol}}) = \frac{\lambda^2}{4\pi} G_{\mathrm{pa\,MAX}}(\mathbf{u}_{\mathrm{pol}}) \,; \tag{15}$$

$$A_{\mathrm{a}} = \frac{\lambda^2}{4\pi} G_{\mathrm{a\,MAX}} \,; \tag{16}$$

$$A_{\mathrm{aeq\,MEA}} = \frac{N\lambda^2}{8\pi} G_{\mathrm{a\,MEA}} \,; \tag{17}$$

and

$$(N = 2) \Longrightarrow \left( A_{\mathrm{aeq\,MIN}} = \frac{\lambda^2}{4\pi} G_{\mathrm{a\,MIN}} \right) . \tag{18}$$

If the MAA is reciprocal, $\mathbf{Z}_{\mathrm{G}} = \mathbf{Z}_{\mathrm{L}}$ and $\mathbf{Z}_{\mathrm{G}}$ is symmetric, then we can assert that, in a given direction:

$$A_{\mathrm{pr}}(\mathbf{u}_{\mathrm{pol}}) = \frac{\lambda^2}{4\pi} G_{\mathrm{pr\,MAX}}(\mathbf{u}_{\mathrm{pol}}) \,; \tag{19}$$

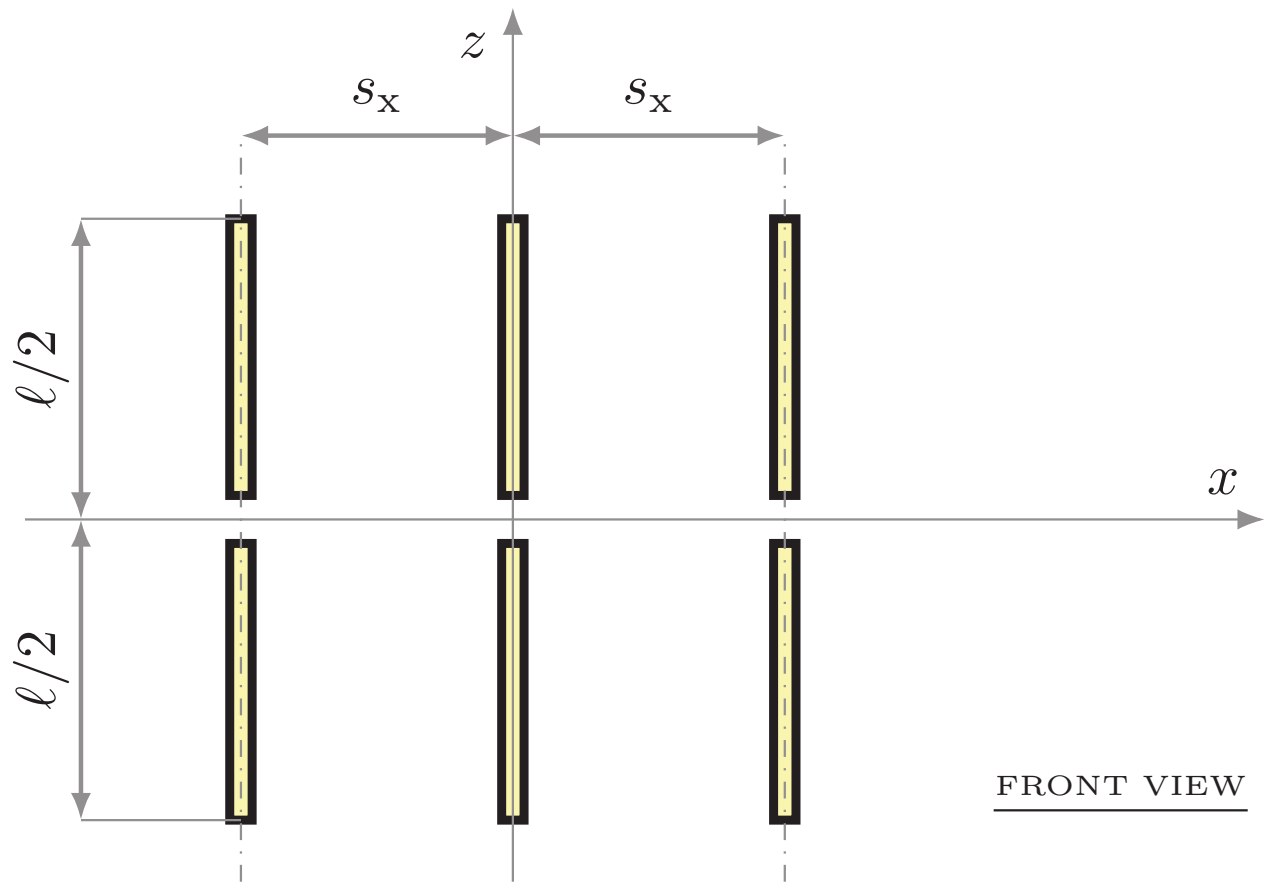

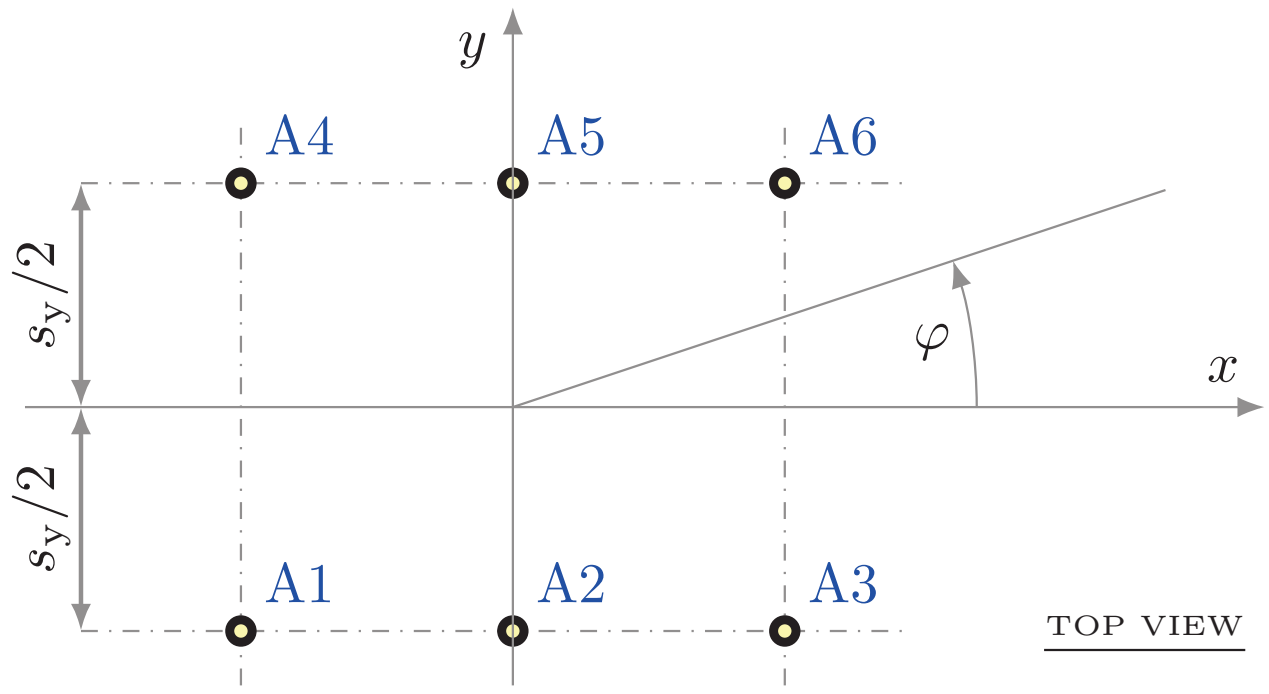

**FIGURE 1.** Multiview drawing (not to scale) of the MAAs considered in Section V. The six identical dipole antennas are labeled “A1” to “A6”.

$$A_{\mathrm{r}} = \frac{\lambda^2}{4\pi} G_{\mathrm{r\,MAX}} \,; \tag{20}$$

$$A_{\mathrm{req\,MEA}} = \frac{N\lambda^2}{8\pi} G_{\mathrm{r\,MEA}} \,; \tag{21}$$

and

$$(N = 2) \Longrightarrow \left( A_{\mathrm{req\,MIN}} = \frac{\lambda^2}{4\pi} G_{\mathrm{r\,MIN}} \right) . \tag{22}$$

The formulas (15)–(22) prove the relevance of the definitions of Section III to the characterization of a MAA for emission and reception. Moreover, (16) and (20) have been used in [2, Sec. XIII–XIV] to obtain: new generalizations of the Friis transmission formula, which are quite different from the ones proposed in [6, Sec. XIII], [9], [10] and [11]; and other new transmission formulas.

## V. EXAMPLES

A 6-port antenna array configuration, made of six perfectly conducting vertical center-fed cylindrical dipole antennas lying in free space, is shown in the not-to-scale drawing of Fig. 1. The total length of each antenna is $\ell = 0.94\,\lambda/2$. All antennas have the same wire diameter $\ell/50$. Between adjacent antennas, the spacing is $s_{\mathrm{x}}$ in the $x$ direction and $s_{\mathrm{y}}$ in the $y$ direction.

We consider two MAAs: in the first one, $s_{\mathrm{x}} = \lambda/4$ and $s_{\mathrm{y}} = \lambda/2$; and in the second one, $s_{\mathrm{x}} = s_{\mathrm{y}} = \lambda/4$.

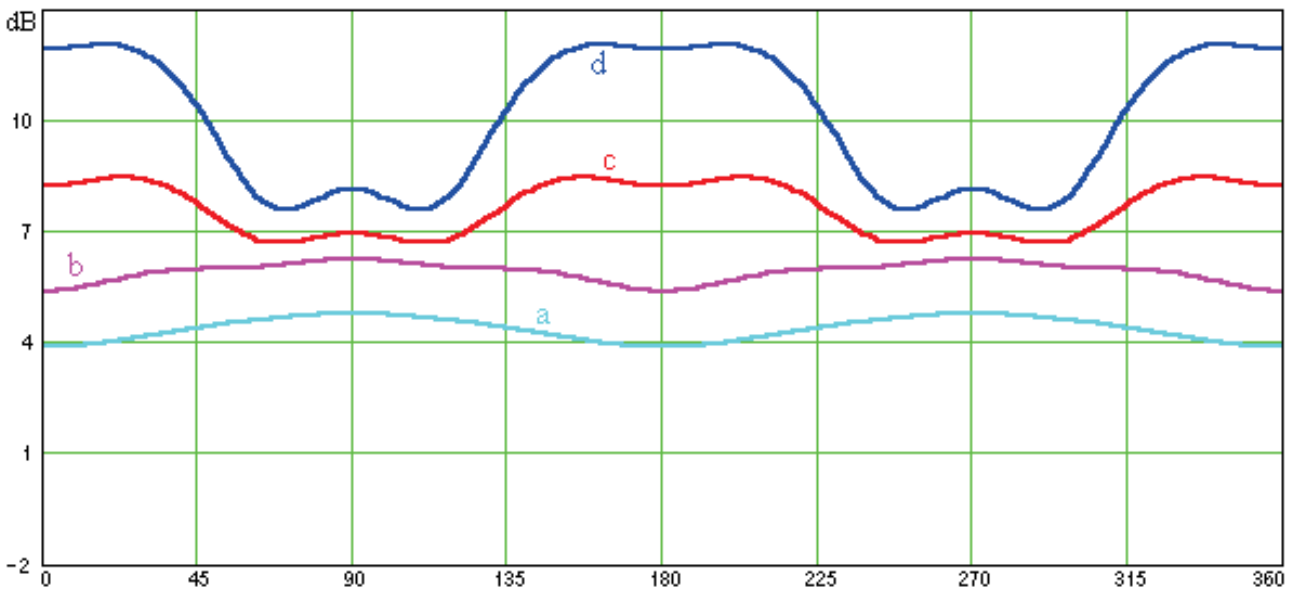

**FIGURE 2.** $G_{\mathrm{a\,MAX}}$ for the MAA in which $s_{\mathrm{y}} = \lambda/2$, as a function of $\varphi$. Curve "a" is for $\theta = 30°$ or $\theta = 150°$. Curve "b" is for $\theta = 45°$ or $\theta = 135°$. Curve "c" is for $\theta = 60°$ or $\theta = 120°$. Curve "d" is for $\theta = 90°$.

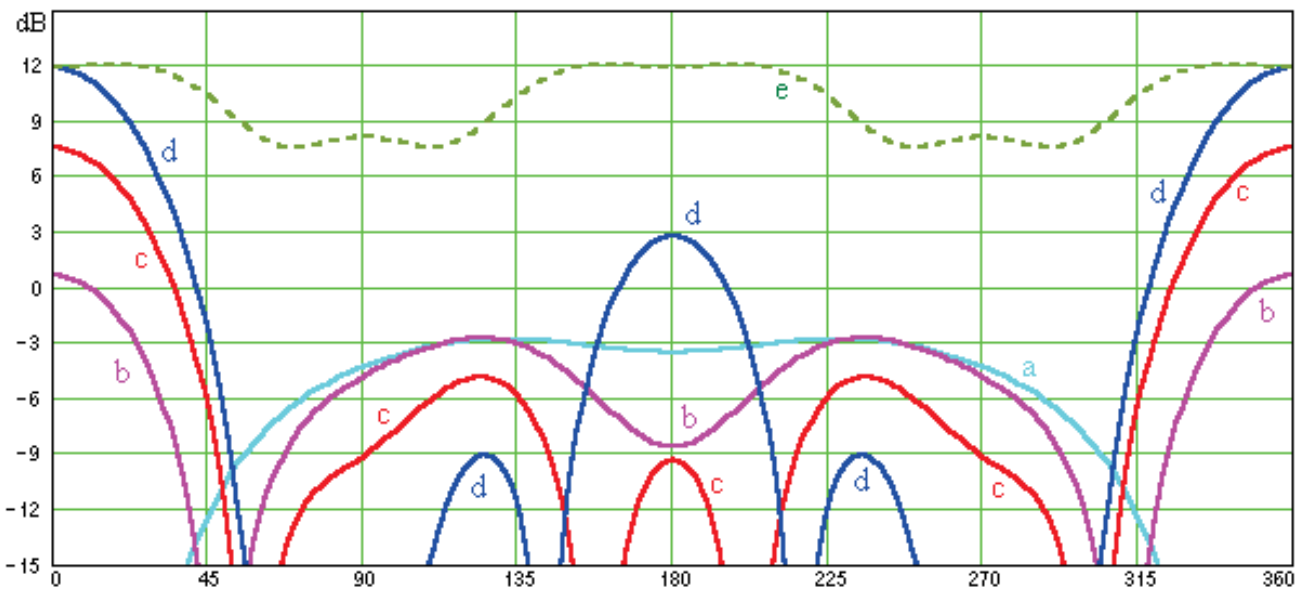

**FIGURE 3.** $G_{\mathrm{a}}$ as a function of $\varphi$ for $s_{\mathrm{y}} = \lambda/2$ and an excitation that maximizes $G_{\mathrm{a}}$ in the direction $\theta = 90°$ and $\varphi = 0°$. Curve "a" is for $\theta = 30°$ or $\theta = 150°$. Curve "b" is for $\theta = 45°$ or $\theta = 135°$. Curve "c" is for $\theta = 60°$ or $\theta = 120°$. Curve "d" is for $\theta = 90°$. Curve "e" is $G_{\mathrm{a\,MAX}}$ for $\theta = 90°$.

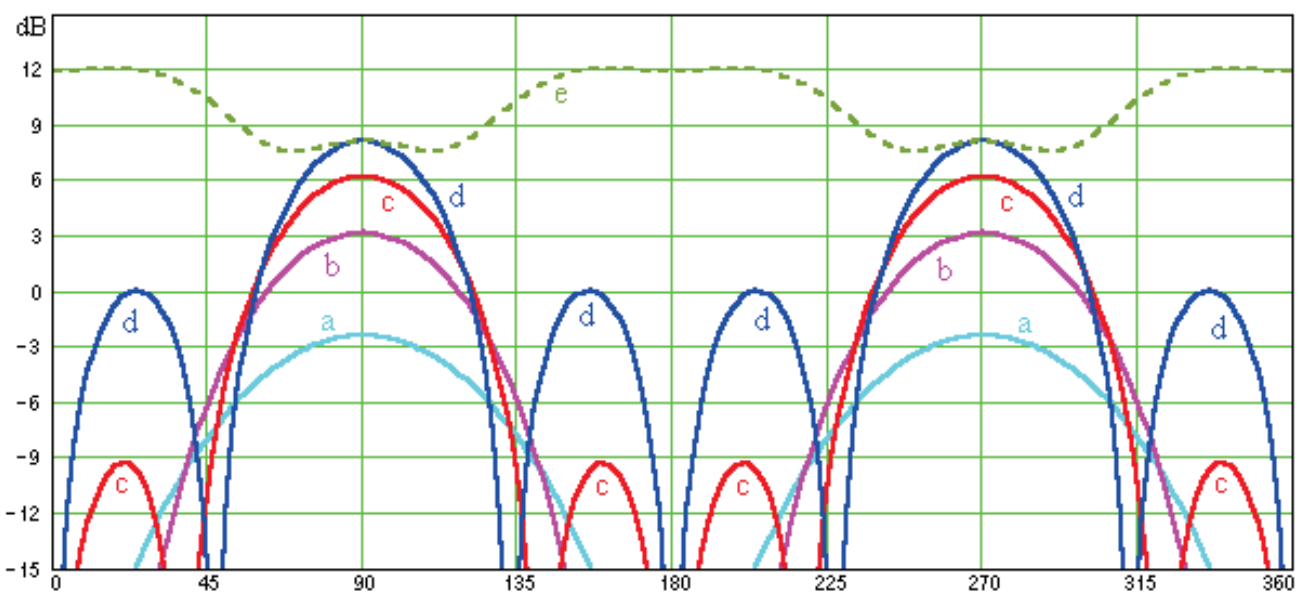

**FIGURE 4.** $G_{\mathrm{a}}$ as a function of $\varphi$ for $s_{\mathrm{y}} = \lambda/2$ and an excitation that maximizes $G_{\mathrm{a}}$ in the direction $\theta = 90°$ and $\varphi = 90°$. Curve "a" is for $\theta = 30°$ or $\theta = 150°$. Curve "b" is for $\theta = 45°$ or $\theta = 135°$. Curve "c" is for $\theta = 60°$ or $\theta = 120°$. Curve "d" is for $\theta = 90°$. Curve "e" is $G_{\mathrm{a\,MAX}}$ for $\theta = 90°$.

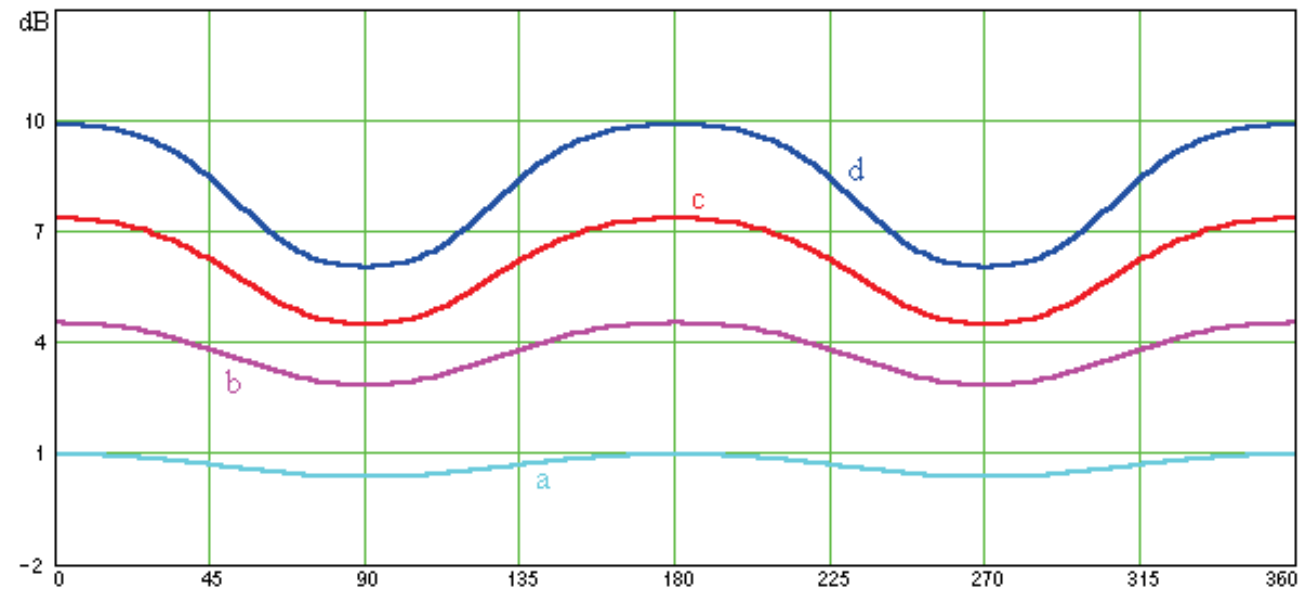

**FIGURE 5.** $G_{\mathrm{r\,MAX}}$ for the MAA in which $s_{\mathrm{y}} = \lambda/2$, as a function of $\varphi$. Curve "a" is for $\theta = 30°$ or $\theta = 150°$. Curve "b" is for $\theta = 45°$ or $\theta = 135°$. Curve "c" is for $\theta = 60°$ or $\theta = 120°$. Curve "d" is for $\theta = 90°$.

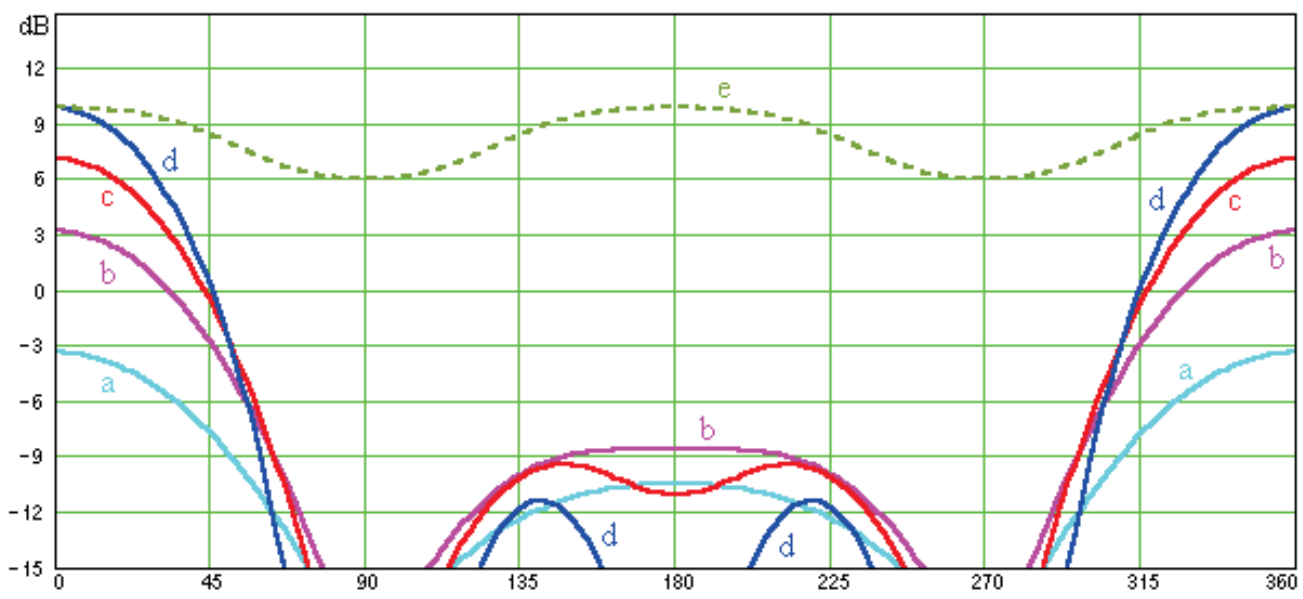

**FIGURE 6.** $G_{\mathrm{r}}$ as a function of $\varphi$ for $s_{\mathrm{y}} = \lambda/2$ and an excitation that maximizes $G_{\mathrm{r}}$ in the direction $\theta = 90°$ and $\varphi = 0°$. Curve "a" is for $\theta = 30°$ or $\theta = 150°$. Curve "b" is for $\theta = 45°$ or $\theta = 135°$. Curve "c" is for $\theta = 60°$ or $\theta = 120°$. Curve "d" is for $\theta = 90°$. Curve "e" is $G_{\mathrm{r\,MAX}}$ for $\theta = 90°$.

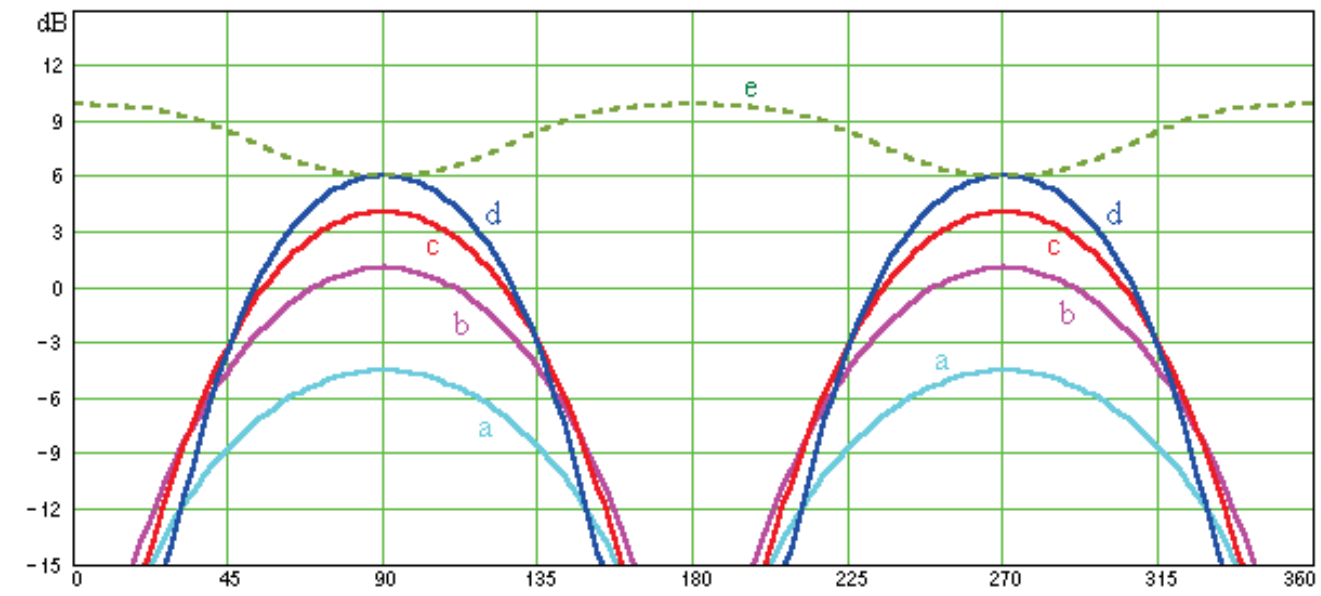

**FIGURE 7.** $G_{\mathrm{r}}$ as a function of $\varphi$ for $s_{\mathrm{y}} = \lambda/2$ and an excitation that maximizes $G_{\mathrm{r}}$ in the direction $\theta = 90°$ and $\varphi = 90°$. Curve "a" is for $\theta = 30°$ or $\theta = 150°$. Curve "b" is for $\theta = 45°$ or $\theta = 135°$. Curve "c" is for $\theta = 60°$ or $\theta = 120°$. Curve "d" is for $\theta = 90°$. Curve "e" is $G_{\mathrm{r\,MAX}}$ for $\theta = 90°$.

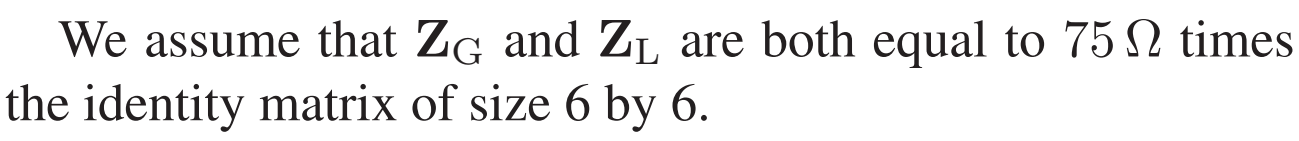

We assume that $\mathbf{Z}_{\mathrm{G}}$ and $\mathbf{Z}_{\mathrm{L}}$ are both equal to $75\,\Omega$ times the identity matrix of size 6 by 6.

Let $(r, \theta, \varphi)$ be the spherical coordinates system associated with the cartesian coordinate system $(x, y, z)$.

For the first MAA, $G_{\mathrm{a\,MAX}}$ is shown in Fig. 2, and $G_{\mathrm{r\,MAX}}$ is shown in Fig. 5, as a function of $\varphi$, for different values of $\theta$. We have to keep in mind that $G_{\mathrm{a\,MAX}}$ is a value of $G_{\mathrm{a}}$ determined for an excitation that depends on the values of $\varphi$ and $\theta$. Likewise $G_{\mathrm{r\,MAX}}$ is a value of $G_{\mathrm{r}}$ determined for an excitation that depends on the value of $\varphi$ and $\theta$.

For the first MAA, $G_{\mathrm{a}}$ is shown in Fig. 3 for an excitation that maximizes $G_{\mathrm{a}}$ in the direction $\theta = 90°$ and $\varphi = 0°$, and in Fig. 4 for an excitation that maximizes $G_{\mathrm{a}}$ in the direction $\theta = 90°$ and $\varphi = 90°$. Thus, in the curves (a) to (d) of Fig. 3, the same excitation has been used for all values of $\varphi$ and $\theta$, and this excitation is such that $G_{\mathrm{a}} = G_{\mathrm{a\,MAX}}$ in the direction $\theta = 90°$ and $\varphi = 0°$. Likewise, in the curves (a) to (d) of Fig. 4, the same excitation has been used for all values of $\varphi$ and $\theta$, and this excitation is such that $G_{\mathrm{a}} = G_{\mathrm{a\,MAX}}$ in the direction $\theta = 90°$ and $\varphi = 90°$. We see that each curve representing $G_{\mathrm{a\,MAX}}$ for a given value of $\theta$ as a function of $\varphi$ in Fig. 2 is an envelope of the curves that would, for all possible excitations, represent $G_{\mathrm{a}}$ for this value of $\theta$ as a function of $\varphi$. Furthermore, we also see that the surface that would represent $G_{\mathrm{a\,MAX}}$ as a function of $\theta$ and $\varphi$ would be an envelope of the surfaces that would, for all possible excitations, represent $G_{\mathrm{a}}$ as a function of $\theta$ and $\varphi$.

For the first MAA, $G_{\mathrm{r}}$ is shown in Fig. 6 for an excitation that maximizes $G_{\mathrm{r}}$ in the direction $\theta = 90°$ and $\varphi = 0°$, and in Fig. 7 for an excitation that maximizes $G_{\mathrm{r}}$ in the direction $\theta = 90°$ and $\varphi = 90°$. Thus, in the curves (a) to (d) of Fig. 6, the same excitation has been used for all values of $\varphi$ and $\theta$,

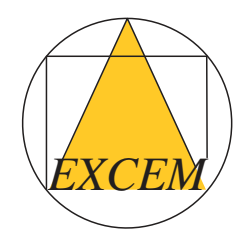

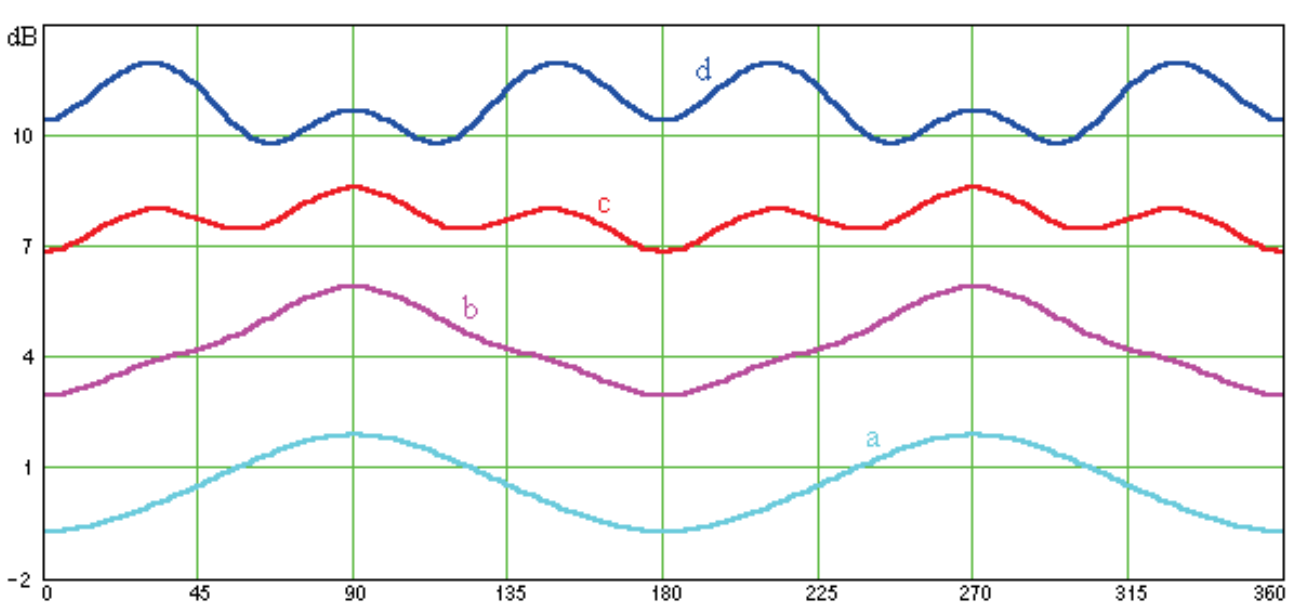

**FIGURE 8.** $G_{\mathrm{a\,MAX}}$ for the MAA in which $s_{\mathrm{y}} = \lambda/4$, as a function of $\varphi$. Curve "a" is for $\theta = 30°$ or $\theta = 150°$. Curve "b" is for $\theta = 45°$ or $\theta = 135°$. Curve "c" is for $\theta = 60°$ or $\theta = 120°$. Curve "d" is for $\theta = 90°$.

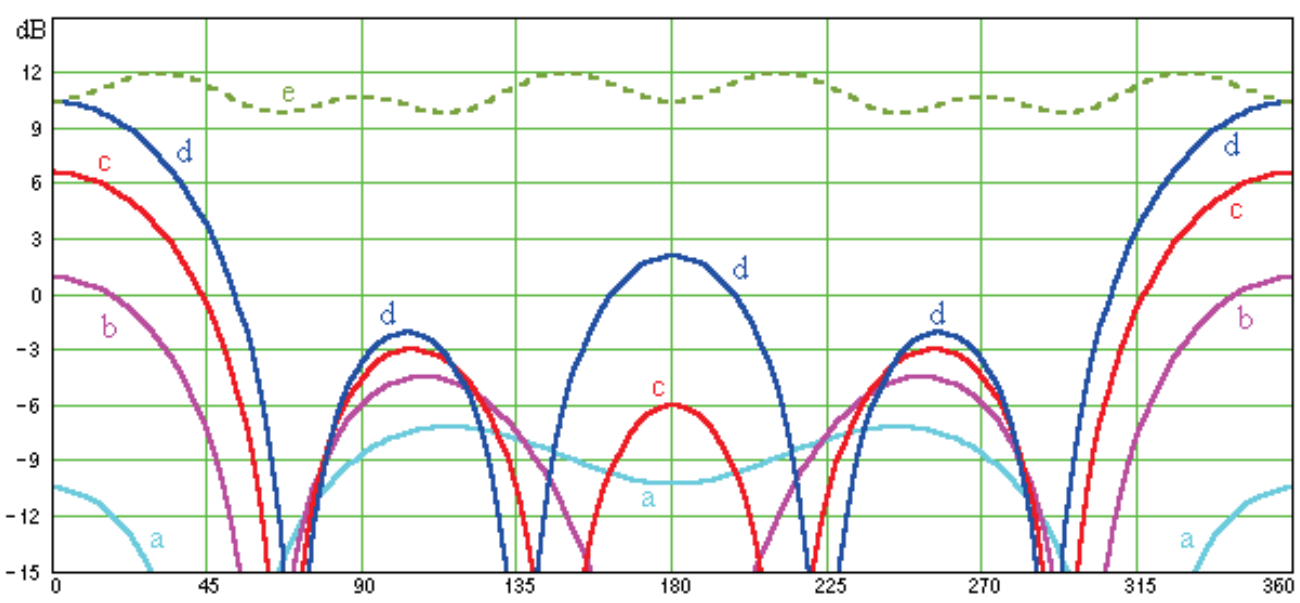

**FIGURE 9.** $G_{\mathrm{a}}$ as a function of $\varphi$ for $s_{\mathrm{y}} = \lambda/4$ and an excitation that maximizes $G_{\mathrm{a}}$ in the direction $\theta = 90°$ and $\varphi = 0°$. Curve "a" is for $\theta = 30°$ or $\theta = 150°$. Curve "b" is for $\theta = 45°$ or $\theta = 135°$. Curve "c" is for $\theta = 60°$ or $\theta = 120°$. Curve "d" is for $\theta = 90°$. Curve "e" is $G_{\mathrm{a\,MAX}}$ for $\theta = 90°$.

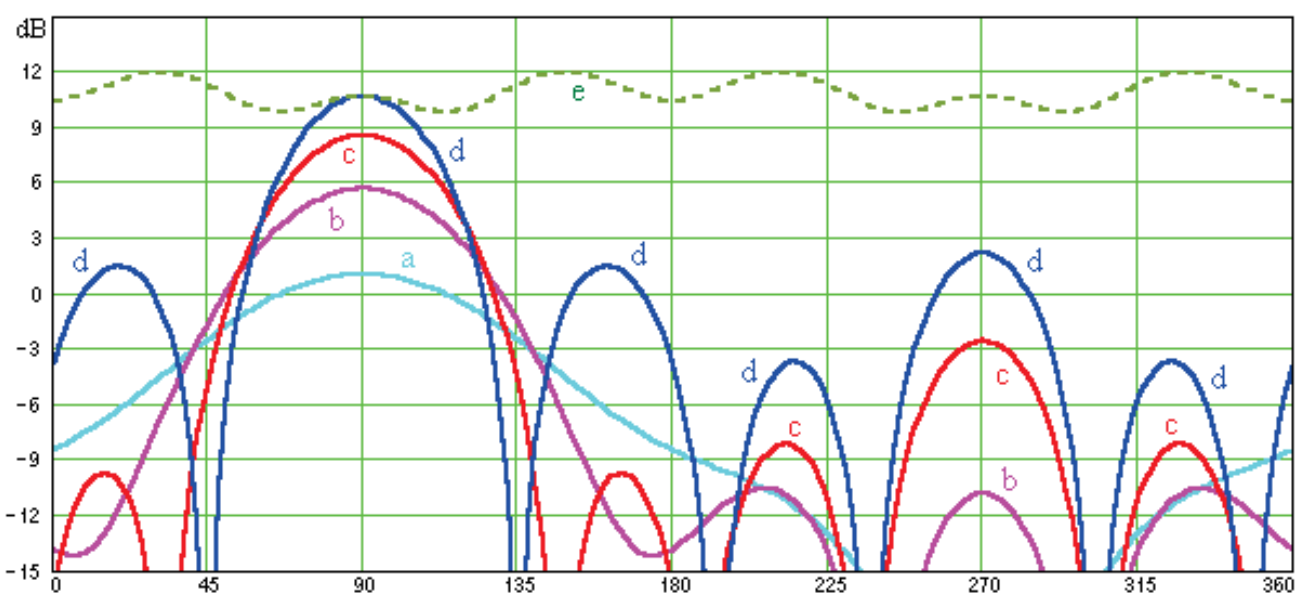

**FIGURE 10.** $G_{\mathrm{a}}$ as a function of $\varphi$ for $s_{\mathrm{y}} = \lambda/4$ and an excitation that maximizes $G_{\mathrm{a}}$ in the direction $\theta = 90°$ and $\varphi = 90°$. Curve "a" is for $\theta = 30°$ or $\theta = 150°$. Curve "b" is for $\theta = 45°$ or $\theta = 135°$. Curve "c" is for $\theta = 60°$ or $\theta = 120°$. Curve "d" is for $\theta = 90°$. Curve "e" is $G_{\mathrm{a\,MAX}}$ for $\theta = 90°$.

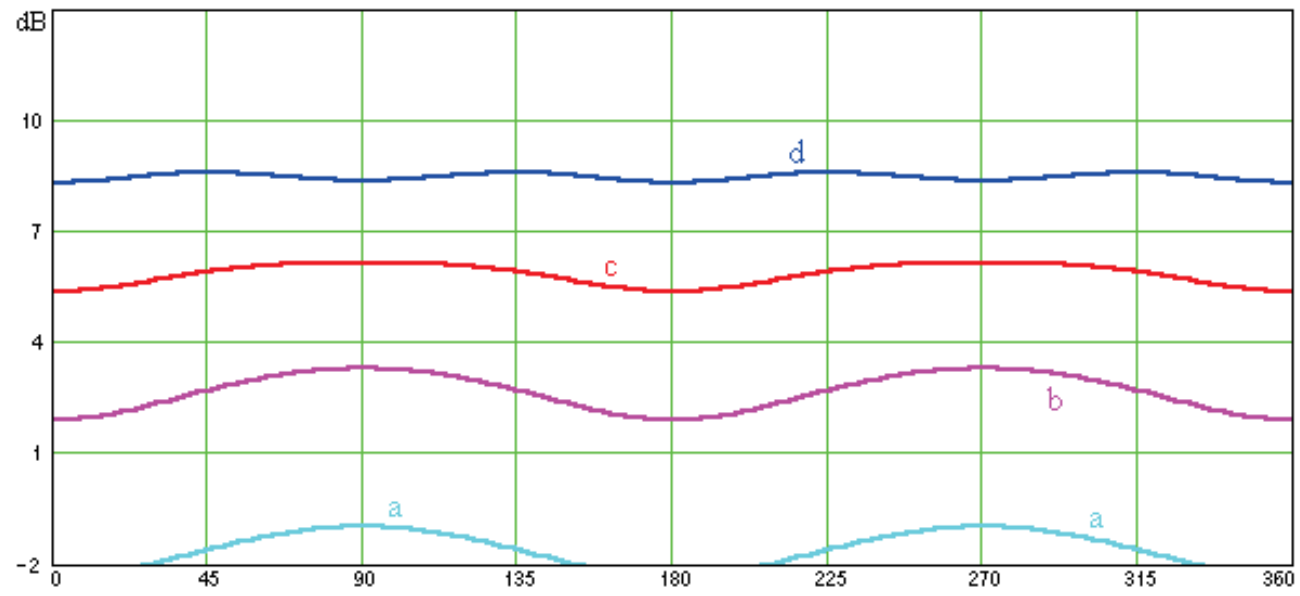

**FIGURE 11.** $G_{\mathrm{r\,MAX}}$ for the MAA in which $s_{\mathrm{y}} = \lambda/4$, as a function of $\varphi$. Curve "a" is for $\theta = 30°$ or $\theta = 150°$. Curve "b" is for $\theta = 45°$ or $\theta = 135°$. Curve "c" is for $\theta = 60°$ or $\theta = 120°$. Curve "d" is for $\theta = 90°$.

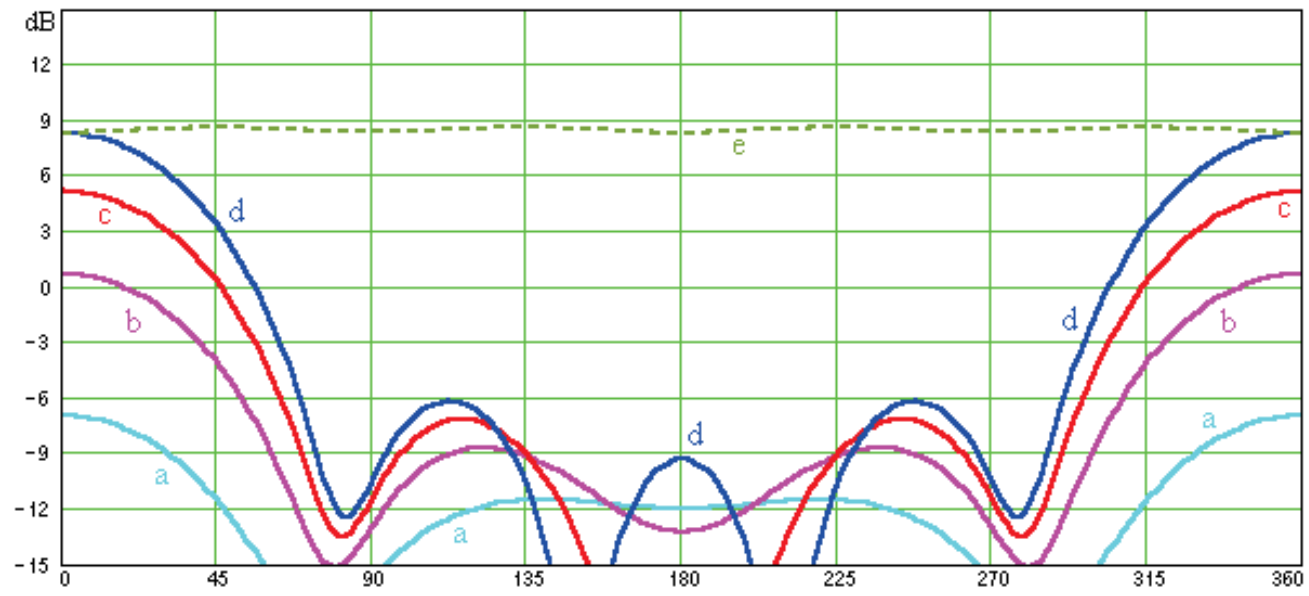

**FIGURE 12.** $G_{\mathrm{r}}$ as a function of $\varphi$ for $s_{\mathrm{y}} = \lambda/4$ and an excitation that maximizes $G_{\mathrm{r}}$ in the direction $\theta = 90°$ and $\varphi = 0°$. Curve "a" is for $\theta = 30°$ or $\theta = 150°$. Curve "b" is for $\theta = 45°$ or $\theta = 135°$. Curve "c" is for $\theta = 60°$ or $\theta = 120°$. Curve "d" is for $\theta = 90°$. Curve "e" is $G_{\mathrm{r\,MAX}}$ for $\theta = 90°$.

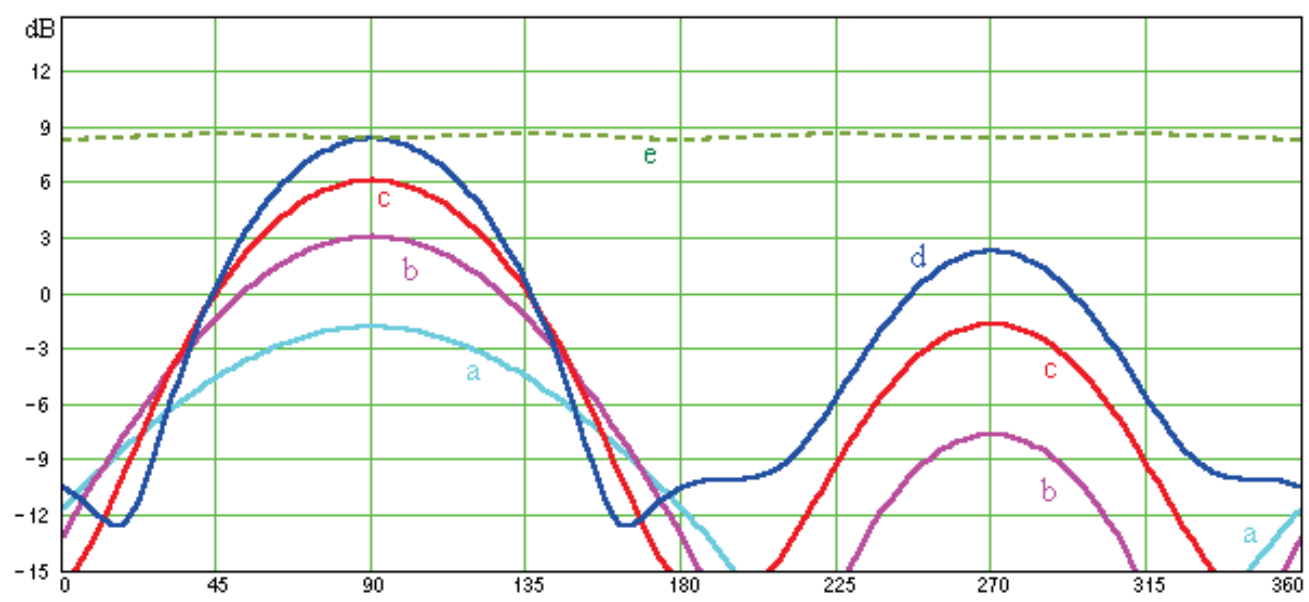

**FIGURE 13.** $G_{\mathrm{r}}$ as a function of $\varphi$ for $s_{\mathrm{y}} = \lambda/4$ and an excitation that maximizes $G_{\mathrm{r}}$ in the direction $\theta = 90°$ and $\varphi = 90°$. Curve "a" is for $\theta = 30°$ or $\theta = 150°$. Curve "b" is for $\theta = 45°$ or $\theta = 135°$. Curve "c" is for $\theta = 60°$ or $\theta = 120°$. Curve "d" is for $\theta = 90°$. Curve "e" is $G_{\mathrm{r\,MAX}}$ for $\theta = 90°$.

and this excitation is such that $G_{\mathrm{r}} = G_{\mathrm{r\,MAX}}$ in the direction $\theta = 90°$ and $\varphi = 0°$. Likewise, in the curves (a) to (d) of Fig. 7, the same excitation has been used for all values of $\varphi$ and $\theta$, and this excitation is such that $G_{\mathrm{r}} = G_{\mathrm{r\,MAX}}$ in the direction $\theta = 90°$ and $\varphi = 90°$. We see that each curve representing $G_{\mathrm{r\,MAX}}$ for a given value of $\theta$ as a function of $\varphi$ in Fig. 5 is an envelope of the curves that would, for all possible excitations, represent $G_{\mathrm{r}}$ for this value of $\theta$ as a function of $\varphi$. Furthermore, we also see that the surface that would represent $G_{\mathrm{r\,MAX}}$ as a function of $\theta$ and $\varphi$ would be an envelope of the surfaces that would, for all possible excitations, represent $G_{\mathrm{r}}$ as a function of $\theta$ and $\varphi$.

Interestingly, the shapes of the curves of Fig 5 to Fig. 7 showing $G_{\mathrm{r\,MAX}}$ and $G_{\mathrm{r}}$ of the first MAA are significantly different from the shapes of the corresponding curves of Fig 2 to Fig. 4 showing $G_{\mathrm{a\,MAX}}$ and $G_{\mathrm{a}}$ of the same MAA. The reason is that the ratio of $G_{\mathrm{r}}$ to $G_{\mathrm{a}}$ is a power transfer ratio, which depends on the excitation [12]. If this power transfer ratio did not depend on the excitation, the corresponding curves would have the same shape.

Fig. 8 to Fig. 13 are about the second MAA, and they correspond to Fig. 2 to Fig. 7 relating to the first MAA, respectively. The above comments on Fig. 2 to Fig. 7 are applicable, mutatis mutandis, to Fig. 8 to Fig. 13, respectively.

Fig. 2 to Fig. 13 allow a quick comparison of some characteristics of the MAAs. For instance, we see that the variations of $G_{\mathrm{a\,MAX}}$ and $G_{\mathrm{r\,MAX}}$ as a function of $\varphi$ are larger for the first MAA than for the second one. This seems related to the fact that, for the first MAA and an excitation that maximizes $G_{\mathrm{a}}$ or $G_{\mathrm{r}}$ in the direction $\theta = 90°$ and $\varphi = 90°$, $G_{\mathrm{a}}$ or $G_{\mathrm{r}}$, respectively, has two equal lobes in the horizontal plane $\theta = 90°$, shown in Fig. 4 and Fig. 7, respectively.

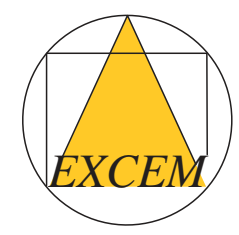

Some information on the computation technique used to obtain Fig. 2 to Fig. 13 is provided in Appendix B.

General results, applicable to both MAAs and others similarly made of thin vertical cylindrical antennas, are obtained in Appendix C. According to (71)–(78) of Appendix C, for each of these MAAs:

- we have $G_{\mathrm{a\,MIN}} = 0$ and $G_{\mathrm{r\,MIN}} = 0$;
- $G_{\mathrm{a\,MEA}}$, $G_{\mathrm{pa\,MAX}}(\mathbf{u}_{\mathrm{pol}})$ and $G_{\mathrm{pa\,MEA}}(\mathbf{u}_{\mathrm{pol}})$ can be easily computed if $G_{\mathrm{a\,MAX}}$ is known; and
- $G_{\mathrm{r\,MEA}}$, $G_{\mathrm{pr\,MAX}}(\mathbf{u}_{\mathrm{pol}})$ and $G_{\mathrm{pr\,MEA}}(\mathbf{u}_{\mathrm{pol}})$ can be easily computed if $G_{\mathrm{r\,MAX}}$ is known.

Consequently, among the emission parameters listed in Table 2, we have only plotted $G_{\mathrm{a\,MAX}}$ and $G_{\mathrm{r\,MAX}}$.

Also, we have not plotted any one of the reception parameters listed in Table 3, because:

- for the first and second MAAs, and others similarly made of parallel thin vertical cylindrical antennas, we have $A_{\mathrm{aeq\,MIN}} = 0\,\mathrm{m}^2$ and $A_{\mathrm{req\,MIN}} = 0\,\mathrm{m}^2$ according to (86) and (89) of Appendix C; and
- the other reception parameters can be easily computed from the emission parameters, using (15)–(17), (19)–(21) and the facts that the MAAs are reciprocal and $\mathbf{Z}_{\mathrm{G}} = \mathbf{Z}_{\mathrm{L}}$ is symmetric.

## VI. CHOICE OF EXCITATION VARIABLE

Up to now in this article, as in [2], the variable used to define an excitation applied to the MAA during emission is the column vector of the rms currents flowing into ports 1 to $N$ of the MAA, denoted by $\mathbf{I}_{\mathrm{A}}$. As said in above in Section II, $\mathbf{I}_{\mathrm{A}}$ can take on any value lying in $\mathbb{C}^N$, because $\mathbf{Z}_{\mathrm{A}}$ exists.

To define the excitation applied to the MAA during emission, we can also use other excitation variables, for instance any one of the excitation variables used in [7, Sec. IV–V]:

- the column vector of the rms open-circuit voltages at ports 1 to $N$ of the MGA, denoted by $\mathbf{V}_{\mathrm{OG}}$ and such that
$$\mathbf{V}_{\mathrm{OG}} = (\mathbf{Z}_{\mathrm{A}} + \mathbf{Z}_{\mathrm{G}})\mathbf{I}_{\mathrm{A}}\,; \tag{23}$$
- the column vector of the rms short-circuit currents at ports 1 to $N$ of the MGA, denoted by $\mathbf{I}_{\mathrm{SG}}$ and such that
$$\mathbf{I}_{\mathrm{SG}} = (\mathbf{1}_N + \mathbf{Z}_{\mathrm{G}}^{-1}\mathbf{Z}_{\mathrm{A}})\mathbf{I}_{\mathrm{A}}\,, \tag{24}$$
where $\mathbf{1}_N$ denotes the identity matrix of size $N$ by $N$;
- the column vector of the rms voltages at ports 1 to $N$ of the MAA, denoted by $\mathbf{V}_{\mathrm{A}}$ and such that
$$\mathbf{V}_{\mathrm{A}} = \mathbf{Z}_{\mathrm{A}}\mathbf{I}_{\mathrm{A}}\,; \tag{25}$$
- the column vector of the normalized rms incident voltages at ports 1 to $N$ of the MAA, denoted by $\mathbf{a}$ and defined for a specified choice of $N$ arbitrary real and positive reference resistances $r_{01}, \ldots, r_{0N}$, by [13]
$$\mathbf{a} = \mathbf{r}_0^{-1/2}\,\frac{\mathbf{V}_{\mathrm{A}} + \mathbf{r}_0\mathbf{I}_{\mathrm{A}}}{2}\,, \tag{26}$$
where $\mathbf{r}_0$ denotes the diagonal matrix of diagonal entries $r_{01}, \ldots, r_{0N}$, and $\mathbf{r}_0^{-1/2}$ denotes the diagonal matrix of diagonal entries $1/\sqrt{r_{01}}, \ldots, 1/\sqrt{r_{0N}}$; or
- the column vector of the rms power waves incident at ports 1 to $N$ of the MAA, defined for a specified choice of $N$ arbitrary complex impedances $z_{01}, \ldots, z_{0N}$ such that $r_{0p} = \mathrm{Re}(z_{0p})$ is positive for any $p \in \{1, \ldots, N\}$, this column vector being denoted by $\hat{\mathbf{a}}$ and given by [14]
$$\hat{\mathbf{a}} = \mathbf{r}_0^{-1/2}\,\frac{\mathbf{V}_{\mathrm{A}} + \mathbf{z}_0\mathbf{I}_{\mathrm{A}}}{2}\,, \tag{27}$$
where $\mathbf{z}_0$ denotes the diagonal matrix of diagonal entries $z_{01}, \ldots, z_{0N}$, and $\mathbf{r}_0^{-1/2}$ denotes the diagonal matrix of diagonal entries $1/\sqrt{r_{01}}, \ldots, 1/\sqrt{r_{0N}}$.

The fact, mentioned in Section II, that $\mathbf{Z}_{\mathrm{G}}$ is invertible was used to obtain (24). Note that $\hat{\mathbf{a}}$ may also be viewed as the column vector of the rms pseudo-waves incident at ports 1 to $N$ of the MAA [15].

Let $\mathbf{X}$ denote one of the excitation variables $\mathbf{V}_{\mathrm{OG}}$, $\mathbf{I}_{\mathrm{SG}}$, $\mathbf{V}_{\mathrm{A}}$, $\mathbf{I}_{\mathrm{A}}$, $\mathbf{a}$, or $\hat{\mathbf{a}}$. We have
$$\mathbf{X} = \mathcal{C}\,\mathbf{I}_{\mathrm{A}}\,, \tag{28}$$
where $\mathcal{C}$ is a matrix of size $N$ by $N$ that depends on the selected excitation variable $\mathbf{X}$, as shown in Table 4.

**TABLE 4.** Some excitation variables $\mathbf{X}$ and associated matrix $\mathcal{C}$.

| Variable $\mathbf{X}$ | $\mathcal{C}$ |
|---|---|
| $\mathbf{V}_{\mathrm{OG}}$ | $\mathbf{Z}_{\mathrm{A}} + \mathbf{Z}_{\mathrm{G}}$ |
| $\mathbf{I}_{\mathrm{SG}}$ | $\mathbf{Z}_{\mathrm{G}}^{-1}(\mathbf{Z}_{\mathrm{A}} + \mathbf{Z}_{\mathrm{G}})$ |
| $\mathbf{V}_{\mathrm{A}}$ | $\mathbf{Z}_{\mathrm{A}}$ |
| $\mathbf{I}_{\mathrm{A}}$ | $\mathbf{1}_N$ |
| $\mathbf{a}$ | $\mathbf{r}_0^{-1/2}(\mathbf{Z}_{\mathrm{A}} + \mathbf{r}_0)/2$ |
| $\hat{\mathbf{a}}$ | $\mathbf{r}_0^{-1/2}(\mathbf{Z}_{\mathrm{A}} + \mathbf{z}_0)/2$ |

In Section II, we observed that $\mathbf{Z}_{\mathrm{A}}+\mathbf{Z}_{\mathrm{G}}$ is invertible. Since we assumed that $H(\mathbf{Z}_{\mathrm{A}})$ is positive definite, $H(\mathbf{Z}_{\mathrm{A}} + \mathbf{r}_0)$ and $H(\mathbf{Z}_{\mathrm{A}} + \mathbf{z}_0)$ are also positive definite, and it follows from Lemma 1 of [16] that $\mathbf{Z}_{\mathrm{A}}$, $\mathbf{Z}_{\mathrm{A}} + \mathbf{r}_0$ and $\mathbf{Z}_{\mathrm{A}} + \mathbf{z}_0$ are invertible. Consequently, it follows from our assumptions that all matrices $\mathcal{C}$ shown in Table 4 are invertible. Thus, any one of the excitation variables $\mathbf{V}_{\mathrm{OG}}$, $\mathbf{I}_{\mathrm{SG}}$, $\mathbf{V}_{\mathrm{A}}$, $\mathbf{I}_{\mathrm{A}}$, $\mathbf{a}$, or $\hat{\mathbf{a}}$ can take on any value lying in $\mathbb{C}^N$.

More generally, we now use $\mathbf{X}$ to denote an arbitrary excitation variable that satisfies (28) in which $\mathcal{C}$ is an invertible matrix of size $N$ by $N$ defining a change of excitation variable from $\mathbf{I}_{\mathrm{A}}$ to $\mathbf{X}$. Since $\mathcal{C}$ is assumed to be invertible, the excitation variable $\mathbf{X}$ can take on any value lying in $\mathbb{C}^N$. Table 4 is now a non-limiting list of possible excitation variables $\mathbf{X}$, showing the associated matrix $\mathcal{C}$ defining the change of variable from $\mathbf{I}_{\mathrm{A}}$ to $\mathbf{X}$.

By [2, Eq. (8)] and (28), using the arbitrary excitation variable $\mathbf{X}$ and the associated change of variable matrix $\mathcal{C}$, the average power received by the $N$ ports of the MAA is
$$P_{\mathrm{RPA}} = \mathbf{X}^*\,\mathcal{D}_{\mathrm{RPA}}\,\mathbf{X}\,, \tag{29}$$
where we have used the positive definite matrix
$$\mathcal{D}_{\mathrm{RPA}} = \mathcal{C}^{-1*}\,H(\mathbf{Z}_{\mathrm{A}})\,\mathcal{C}^{-1}. \tag{30}$$

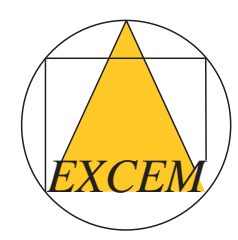

In the special case where $\mathbf{X} = \mathbf{a}$, we also have [13]

$$\mathcal{D}_{\mathrm{RPA}} = \mathbf{1}_N - \mathbf{S}_A^* \mathbf{S}_A \,, \tag{31}$$

where $\mathbf{S}_A$ is the scattering matrix of the MAA for the reference resistances $r_{01}, \ldots, r_{0N}$.

By [2, Eq. (6)–(7)] and (28), using the arbitrary excitation variable $\mathbf{X}$ and the associated change of variable matrix $\mathcal{C}$, the available power of the MGA is

$$P_{\mathrm{AVG}} = \mathbf{X}^* \, \mathcal{D}_{\mathrm{AVG}} \, \mathbf{X} \,, \tag{32}$$

where we have used the positive definite matrix

$$\mathcal{D}_{\mathrm{AVG}} = \mathcal{C}^{-1*} \, \mathbf{Z}_{\mathrm{AVG}} \, \mathcal{C}^{-1}. \tag{33}$$

In the special case where $\mathbf{X} = \mathbf{V}_{\mathrm{OG}}$, it follows from (3) and (33) that

$$\mathcal{D}_{\mathrm{AVG}} = \frac{1}{4} \, H(\mathbf{Z}_{\mathrm{G}})^{-1} \,, \tag{34}$$

which does not depend on the MAA.

In the special case where $\mathbf{X} = \mathbf{I}_{\mathrm{SG}}$, it follows from (3) and (33) that

$$\mathcal{D}_{\mathrm{AVG}} = \frac{1}{4} \, H(\mathbf{Y}_{\mathrm{G}})^{-1} \,, \tag{35}$$

which does not depend on the MAA.

In the very special case where $\mathbf{X} = \mathbf{a}$ and $\mathbf{Z}_{\mathrm{G}} = \mathbf{r}_0$, it follows from (3) and (33) that

$$\mathcal{D}_{\mathrm{AVG}} = \mathbf{1}_N \,, \tag{36}$$

which does not depend on the MAA.

In a given direction, using the arbitrary excitation variable $\mathbf{X}$ and the associated change of variable matrix $\mathcal{C}$, it follows from (6)–(9), (28), (30) and (33) that, if $\mathbf{X} \neq \mathbf{0}$, we have

$$G_{\mathrm{pa}}(\mathbf{u}_{\mathrm{pol}}) = \frac{\mathbf{X}^* \, \mathcal{N}_{\mathrm{Ap}}(\mathbf{u}_{\mathrm{pol}}) \, \mathbf{X}}{\mathbf{X}^* \, \mathcal{D}_{\mathrm{RPA}} \, \mathbf{X}} \,, \tag{37}$$

$$G_{\mathrm{a}} = \frac{\mathbf{X}^* \, \mathcal{N}_{\mathrm{A}} \, \mathbf{X}}{\mathbf{X}^* \, \mathcal{D}_{\mathrm{RPA}} \, \mathbf{X}} \,, \tag{38}$$

$$G_{\mathrm{pr}}(\mathbf{u}_{\mathrm{pol}}) = \frac{\mathbf{X}^* \, \mathcal{N}_{\mathrm{Ap}}(\mathbf{u}_{\mathrm{pol}}) \, \mathbf{X}}{\mathbf{X}^* \, \mathcal{D}_{\mathrm{AVG}} \, \mathbf{X}} \,, \tag{39}$$

$$G_{\mathrm{r}} = \frac{\mathbf{X}^* \, \mathcal{N}_{\mathrm{A}} \, \mathbf{X}}{\mathbf{X}^* \, \mathcal{D}_{\mathrm{AVG}} \, \mathbf{X}} \,, \tag{40}$$

where we have used the positive definite matrices

$$\mathcal{N}_{\mathrm{Ap}}(\mathbf{u}_{\mathrm{pol}}) = \mathcal{C}^{-1*} \, \mathbf{N}_{\mathrm{Ap}}(\mathbf{u}_{\mathrm{pol}}) \, \mathcal{C}^{-1} \tag{41}$$

and

$$\mathcal{N}_{\mathrm{A}} = \mathcal{C}^{-1*} \, \mathbf{N}_{\mathrm{A}} \, \mathcal{C}^{-1}, \tag{42}$$

and where $\mathbf{u}_{\mathrm{pol}}$ is a given polarization vector used to specify a given polarization. This allows us to assert that:

- since $\mathcal{N}_{\mathrm{Ap}}(\mathbf{u}_{\mathrm{pol}})\mathcal{D}_{\mathrm{RPA}}^{-1}$ and $\mathbf{N}_{\mathrm{Ap}}(\mathbf{u}_{\mathrm{pol}})H(\mathbf{Z}_{\mathrm{A}})^{-1}$ are similar, they have the same eigenvalues and the same trace, so that $G_{\mathrm{pa\,MAX}}(\mathbf{u}_{\mathrm{pol}})$ and $G_{\mathrm{pa\,MEA}}(\mathbf{u}_{\mathrm{pol}})$ can be obtained from $\mathcal{N}_{\mathrm{Ap}}(\mathbf{u}_{\mathrm{pol}})\mathcal{D}_{\mathrm{RPA}}^{-1}$;
- since $\mathcal{N}_{\mathrm{A}}\mathcal{D}_{\mathrm{RPA}}^{-1}$ and $\mathbf{N}_{\mathrm{A}} H(\mathbf{Z}_{\mathrm{A}})^{-1}$ are similar, these matrices have the same eigenvalues and the same trace, so that $G_{\mathrm{a\,MAX}}$, $G_{\mathrm{a\,MEA}}$ and $G_{\mathrm{a\,MIN}}$ can be obtained from $\mathcal{N}_{\mathrm{A}}\mathcal{D}_{\mathrm{RPA}}^{-1}$;
- since $\mathcal{N}_{\mathrm{Ap}}(\mathbf{u}_{\mathrm{pol}})\mathcal{D}_{\mathrm{AVG}}^{-1}$ and $\mathbf{N}_{\mathrm{Ap}}(\mathbf{u}_{\mathrm{pol}})\mathbf{Z}_{\mathrm{AVG}}^{-1}$ are similar, they have the same eigenvalues and the same trace, so that $G_{\mathrm{pr\,MAX}}(\mathbf{u}_{\mathrm{pol}})$ and $G_{\mathrm{pr\,MEA}}(\mathbf{u}_{\mathrm{pol}})$ can be obtained from $\mathcal{N}_{\mathrm{Ap}}(\mathbf{u}_{\mathrm{pol}})\mathcal{D}_{\mathrm{AVG}}^{-1}$; and
- since $\mathcal{N}_{\mathrm{A}}\mathcal{D}_{\mathrm{AVG}}^{-1}$ and $\mathbf{N}_{\mathrm{A}}\mathbf{Z}_{\mathrm{AVG}}^{-1}$ are similar, they have the same eigenvalues and the same trace, so that $G_{\mathrm{r\,MAX}}$, $G_{\mathrm{r\,MEA}}$ and $G_{\mathrm{r\,MIN}}$ can be obtained from $\mathcal{N}_{\mathrm{A}}\mathcal{D}_{\mathrm{AVG}}^{-1}$.

Thus, we can assert that the choice of the variable used to define the excitation in Section III had no effect on the value of the emission parameters defined in Table 2. This invariance under a change of excitation variable, like the invariance under a change of experimental system considered in [2, Sec. X], is an important property of these emission parameters, that supports their relevance as fundamental characteristics of an MAA.

The invariance under a change of excitation variable does not imply that all possible excitation variables lead to identical computations. For instance, if we want to evaluate the "reached" parameters of Table 1 or Table 2, it might be advantageous to use $\mathbf{X} = \mathbf{V}_{\mathrm{OG}}$ or $\mathbf{X} = \mathbf{I}_{\mathrm{SG}}$, to take advantage of the simplicity of (34)–(35).

The invariance under a change of excitation variable also allows us to compare the emission parameters considered in this article with the ones defined by other authors.

In [17, Sec. 4], a "realized gain" in a given direction is defined for an MAA, using a vector of "incoming waves" as excitation variable. This realized gain might be viewed as a generalization of the realized gain defined in [18] for a single-port antenna. By (36), this realized gain should be identical to our reached gain $G_{\mathrm{r}}$ if $\mathbf{Z}_{\mathrm{G}} = \mathbf{r}_0$, where $\mathbf{r}_0$ is the diagonal matrix of the positive reference resistances used to define the "incoming waves". An "effective area" in a given direction is also defined in [17, Sec. 4]. It does not correspond to the definition of the absolute effective area stated above in Section III.E, and cannot be viewed as a generalization of the classic definition of an effective area for a single-port antenna stated in [3]–[4] because it is not a ratio of an available power at the $N$ ports of the MAA to a power flux density of an incident wave. More precisely, we consider that [17, Eq. (37)] is not correct, so that [17, Eq. (41)] is not correct either. The correct relationships between the emission and reception parameters of a reciprocal MAA are the ones stated above in Section IV.D, which are not consistent with [17, Eq. (41)].

In [19, Sec. VII], a "realized gain" in a given direction is defined for an MAA, and a "maximal realized gain" is derived, using a vector of peak port voltages as excitation variable. By (36), these realized gain and maximal realized gain should be identical to $G_{\mathrm{r}}$ and $G_{\mathrm{r\,MAX}}$, respectively, if we assume $\mathbf{Z}_{\mathrm{G}} = \mathbf{r}_0$, where $\mathbf{r}_0$ is the suitable diagonal matrix of positive reference resistances.

In [20, Sec. III], a partial realized gain in a given direction and along a given complex polarization vector is defined for an MAA, and a maximum partial realized gain is derived, using the vector $\mathbf{a}$ defined above as excitation variable. By (36), these partial realized gain and maximum partial realized gain should be identical to $G_{\mathrm{pr}}(\mathbf{u}_{\mathrm{pol}})$ and $G_{\mathrm{pr\,MAX}}(\mathbf{u}_{\mathrm{pol}})$, respectively, if $\mathbf{Z}_{\mathrm{G}} = \mathbf{r}_0$, where $\mathbf{r}_0$ is defined as in (26).

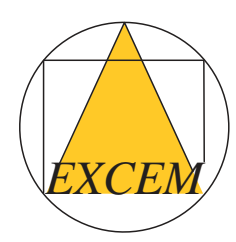

## VII. EFFECT OF A PASSIVE LINEAR EMBEDDING

A passive linear embedding of the ports of the MAA is shown in Fig. 14. It involves a passive and LTI inserted network having 2 sets of ports, referred to as port set 1 and port set 2. Port set 1 consists of $m$ ports numbered from 1 to $m$, where $m$ is a positive integer, and port set 2 consists of $N$ ports numbered from 1 to $N$. When we say that port set 1 is connected to an $m$-port device, we assume that the ports of the $m$-port device are numbered from 1 to $m$, and that, for any integer $p \in \{1, \ldots, m\}$, its port $p$ is connected to port $p$ of port set 1 (positive terminal to positive terminal and negative terminal to negative terminal). Port set 2 is connected to the MAA, that is to say: for any integer $q \in \{1, \ldots, N\}$, port $q$ of the MAA is connected to port $q$ of port set 2 (positive terminal to positive terminal and negative terminal to negative terminal).

The MAA is now referred to as original MAA (OMAA) in this Section VII. The passive linear embedding of the ports of the OMAA creates a new MAA (NMAA) the ports of which are ports 1 to $m$ of port set 1. We want to discuss the emission and reception parameters of the NMAA.

The NMAA has some interesting properties that directly follow from the passivity of the inserted network:

- in a given direction and for a given polarization, the maximum partial absolute gain of the NMAA is less than or equal to the one of the OMAA, and the partial absolute effective area of the NMAA is less than or equal to the one of the OMAA; and
- in a given direction, the maximum absolute gain of the NMAA is less than or equal to the one of the OMAA, the absolute effective area of the NMAA is less than or equal to the one of the OMAA, the mean absolute equivalent area of the NMAA is less than or equal to the one of the OMAA, and the minimum absolute equivalent area of the NMAA is less than or equal to the one of the OMAA.

We can arrange the ports of the inserted network in the following order: port 1 to $m$ of port set 1, and then ports 1 to $N$ of port set 2. Using this order of the ports, we now assume that the inserted network has an impedance matrix, denoted by $\mathbf{Z}_{\mathrm{IN}}$, of size $(m+N)$ by $(m+N)$. It may be partitioned into four submatrices, $\mathbf{Z}_{\mathrm{IN}11}$ of size $m$ by $m$, $\mathbf{Z}_{\mathrm{IN}12}$ of size $m$ by $N$, $\mathbf{Z}_{\mathrm{IN}21}$ of size $N$ by $m$ and $\mathbf{Z}_{\mathrm{IN}22}$ of size $N$ by $N$, which are such that

$$\mathbf{Z}_{\mathrm{IN}} = \begin{pmatrix} \mathbf{Z}_{\mathrm{IN}11} & \mathbf{Z}_{\mathrm{IN}12} \\ \mathbf{Z}_{\mathrm{IN}21} & \mathbf{Z}_{\mathrm{IN}22} \end{pmatrix} . \tag{43}$$

The linear embedding is said to be lossless if and only if the inserted network is lossless. We now assume that the linear embedding is lossless, which is equivalent to $H(\mathbf{Z}_{\mathrm{IN}}) = 0$, and also equivalent to

$$H(\mathbf{Z}_{\mathrm{IN}11}) = 0 , \tag{44}$$

$$H(\mathbf{Z}_{\mathrm{IN}22}) = 0 \tag{45}$$

and

$$\mathbf{Z}_{\mathrm{IN}12} = -\mathbf{Z}^*_{\mathrm{IN}21} . \tag{46}$$

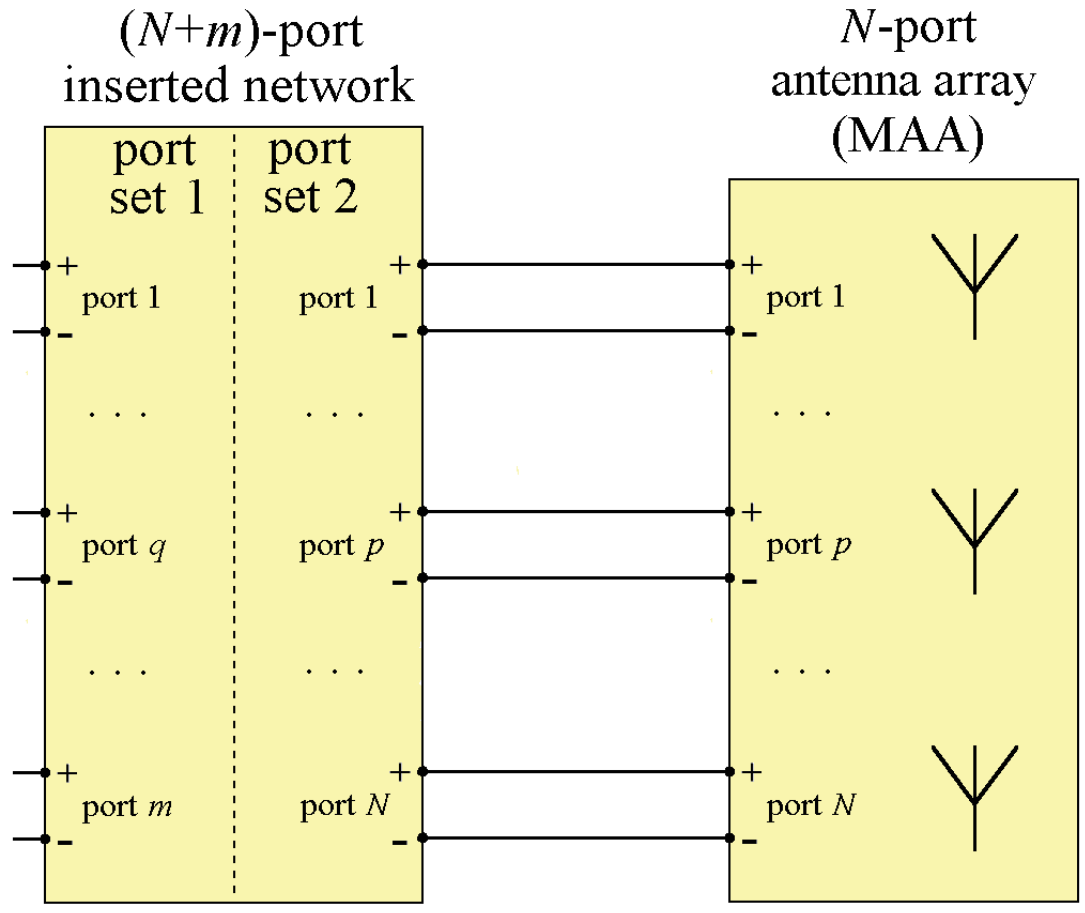

**FIGURE 14.** A passive linear embedding of the MAA ports uses a passive and LTI $(N+m)$-port inserted network to create a new $m$-port MAA.

The "lossless imbeddings" defined and studied in Section 2.2, Fig. 2.2 and Fig. 2.4 of [21] are special cases of the lossless linear embedding of the MAA ports considered here.

$H(\mathbf{Z}_{\mathrm{A}})$ being positive definite, it follows from (45) that $H(\mathbf{Z}_{\mathrm{A}} + \mathbf{Z}_{\mathrm{IN}22})$ is positive definite, and from Lemma 1 of [16] that $\mathbf{Z}_{\mathrm{A}} + \mathbf{Z}_{\mathrm{IN}22}$ is invertible and $H((\mathbf{Z}_{\mathrm{A}} + \mathbf{Z}_{\mathrm{IN}22})^{-1})$ is positive definite. Thus, we find by inspection that the inserted network has a current gain matrix from port set 1 to port set 2, which is given by

$$\mathbf{G}_{\mathrm{IIN}21} = (\mathbf{Z}_{\mathrm{A}} + \mathbf{Z}_{\mathrm{IN}22})^{-1}\, \mathbf{Z}_{\mathrm{IN}21} \tag{47}$$

and satisfies

$$\mathbf{I}_{\mathrm{A}} = \mathbf{G}_{\mathrm{IIN}21}\, \mathbf{I}_{\mathrm{NA}} , \tag{48}$$

where $\mathbf{I}_{\mathrm{NA}}$ is the column vector of the rms currents flowing into ports 1 to $m$ of the NMAA (which are ports 1 to $m$ of port set 1). Hence, we find by inspection that the NMAA has an impedance matrix given by

$$\mathbf{Z}_{\mathrm{NA}} = \mathbf{Z}_{\mathrm{IN}11} - \mathbf{Z}_{\mathrm{IN}12}(\mathbf{Z}_{\mathrm{A}} + \mathbf{Z}_{\mathrm{IN}22})^{-1}\, \mathbf{Z}_{\mathrm{IN}21} . \tag{49}$$

Consequently, $\mathbf{I}_{\mathrm{NA}}$ can take on any value lying in $\mathbb{C}^m$. Using (46) in (49), we find

$$\mathbf{Z}_{\mathrm{NA}} = \mathbf{Z}_{\mathrm{IN}11} + \mathbf{Z}^*_{\mathrm{IN}21}(\mathbf{Z}_{\mathrm{A}} + \mathbf{Z}_{\mathrm{IN}22})^{-1}\, \mathbf{Z}_{\mathrm{IN}21} . \tag{50}$$

Using (44) and (50), we get

$$H(\mathbf{Z}_{\mathrm{NA}}) = \mathbf{Z}^*_{\mathrm{IN}21}\, H\big((\mathbf{Z}_{\mathrm{A}} + \mathbf{Z}_{\mathrm{IN}22})^{-1}\big)\, \mathbf{Z}_{\mathrm{IN}21} . \tag{51}$$

It follows from (51) that $H(\mathbf{Z}_{\mathrm{NA}})$ is positive semidefinite because $H((\mathbf{Z}_{\mathrm{A}} + \mathbf{Z}_{\mathrm{IN}22})^{-1})$ is positive definite. It also follows from (51) and the rank-nullity theorem that, if $\operatorname{rank} \mathbf{Z}_{\mathrm{IN}21} = m$, then $H(\mathbf{Z}_{\mathrm{NA}})$ is positive definite. Thus, the theory presented in Section II to Section VI is fully applicable to the NMAA if $\operatorname{rank} \mathbf{Z}_{\mathrm{IN}21} = m$.

Since the inserted network is lossless, for any $\mathbf{I}_{\mathrm{NA}} \in \mathbb{C}^m$ the average power $\mathbf{I}^*_{\mathrm{NA}} H(\mathbf{Z}_{\mathrm{NA}})\, \mathbf{I}_{\mathrm{NA}}$ received by the NMAA is equal to the average power $\mathbf{I}^*_{\mathrm{NA}} \mathbf{G}^*_{\mathrm{IIN}21} H(\mathbf{Z}_{\mathrm{A}})\, \mathbf{G}_{\mathrm{IIN}21} \mathbf{I}_{\mathrm{NA}}$ received by the OMAA. Accordingly, by [22, Sec. 4.1.P6], we may conclude that

$$H(\mathbf{Z}_{\mathrm{NA}}) = \mathbf{G}^*_{\mathrm{IIN}21} H(\mathbf{Z}_{\mathrm{A}})\, \mathbf{G}_{\mathrm{IIN}21} . \tag{52}$$

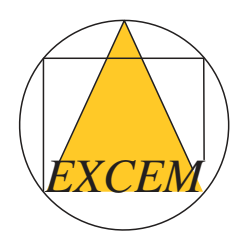

We say that the lossless linear embedding is invertible if and only if $m = N$ and $\operatorname{rank} \mathbf{Z}_{\mathrm{IN21}} = m$. This is because in this case $\mathbf{Z}_{\mathrm{IN21}}$ is square and invertible, hence it follows from (47) that $\mathbf{G}_{\mathrm{IIN21}}$ is also square and invertible.

As regards emission by the NMAA, we assume that port set 1 is connected to an $m$-port generator. If $\operatorname{rank} \mathbf{Z}_{\mathrm{IN21}} = m$, transposing (6)–(7) to the NMAA and using (52), we find that, in a given direction, if $\mathbf{I}_{\mathrm{A}} \neq \mathbf{0}$, then the partial absolute gain of the NMAA for the polarization vector $\mathbf{u}_{\mathrm{pol}}$ is

$$G'_{\mathrm{pa}}(\mathbf{u}_{\mathrm{pol}}) = \frac{\mathbf{I}^*_{\mathrm{NA}} \mathbf{G}^*_{\mathrm{IIN21}} \mathbf{N}_{\mathrm{Ap}}(\mathbf{u}_{\mathrm{pol}}) \mathbf{G}_{\mathrm{IIN21}} \mathbf{I}_{\mathrm{NA}}}{\mathbf{I}^*_{\mathrm{NA}} \mathbf{G}^*_{\mathrm{IIN21}} H(\mathbf{Z}_{\mathrm{A}}) \mathbf{G}_{\mathrm{IIN21}} \mathbf{I}_{\mathrm{NA}}} \tag{53}$$

and the absolute gain of the NMAA is

$$G'_{\mathrm{a}} = \frac{\mathbf{I}^*_{\mathrm{NA}} \mathbf{G}^*_{\mathrm{IIN21}} \mathbf{N}_{\mathrm{A}} \mathbf{G}_{\mathrm{IIN21}} \mathbf{I}_{\mathrm{NA}}}{\mathbf{I}^*_{\mathrm{NA}} \mathbf{G}^*_{\mathrm{IIN21}} H(\mathbf{Z}_{\mathrm{A}}) \mathbf{G}_{\mathrm{IIN21}} \mathbf{I}_{\mathrm{NA}}} \,. \tag{54}$$

If the lossless linear embedding is invertible, $\mathcal{C} = \mathbf{G}^{-1}_{\mathrm{IIN21}}$ defines a change of excitation variable of the OMAA, from $\mathbf{I}_{\mathrm{A}}$ to $\mathbf{I}_{\mathrm{NA}}$. Thus, for the OMAA, (37) and (38) are applicable with $\mathbf{X} = \mathbf{I}_{\mathrm{NA}}$. Using (30) and (41)–(42), we find that, for an invertible lossless linear embedding: the right-hand sides of (37) and (53) are equal; and the right-hand sides of (38) and (54) are equal.

Accordingly, for an invertible lossless linear embedding, we can use the invariance under a change of excitation variable to assert that:

- in a given direction and for a given polarization, the maximum partial absolute gains of the NMAA and OMAA are equal, and the mean partial absolute gains of the NMAA and OMAA are equal; and
- in a given direction, the maximum absolute gains of the NMAA and OMAA are equal, the mean absolute gains of the NMAA and OMAA are equal, and the minimum absolute gains of the NMAA and OMAA are equal.

As regards reception by the NMAA, we assume that port set 1 is connected to an $m$-port load having an impedance matrix $\mathbf{Z}_{\mathrm{LNA}}$ such that $\mathbf{Z}_{\mathrm{LNA}} = \mathbf{Z}^*_{\mathrm{NA}}$, so that this $m$-port load receives the available power of the NMAA because $H(\mathbf{Z}_{\mathrm{NA}})$ is positive semidefinite and [23, Eq. (7)] is applicable in this case. By (50), we have

$$\mathbf{Z}_{\mathrm{LNA}} = \mathbf{Z}^*_{\mathrm{IN11}} + \mathbf{Z}^*_{\mathrm{IN21}} (\mathbf{Z}^*_{\mathrm{A}} + \mathbf{Z}^*_{\mathrm{IN22}})^{-1} \mathbf{Z}_{\mathrm{IN21}} \,. \tag{55}$$

We now assume $\operatorname{rank} \mathbf{Z}_{\mathrm{IN21}} = m$ to ensure that $H(\mathbf{Z}_{\mathrm{NA}})$ is positive definite, so that $H(\mathbf{Z}_{\mathrm{LNA}})$ is positive definite because $\mathbf{Z}_{\mathrm{LNA}} = \mathbf{Z}^*_{\mathrm{NA}}$. Thus, $H(\mathbf{Z}_{\mathrm{LNA}} + \mathbf{Z}_{\mathrm{IN11}})$ is positive definite by (44). It follows from Lemma 1 of [16] that $\mathbf{Z}_{\mathrm{LNA}} + \mathbf{Z}_{\mathrm{IN11}}$ is invertible. Using (46), we find by inspection that the impedance matrix seen by the OMAA is

$$\mathbf{Z}_{\mathrm{L}} = \mathbf{Z}_{\mathrm{IN22}} + \mathbf{Z}_{\mathrm{IN21}} (\mathbf{Z}_{\mathrm{LNA}} + \mathbf{Z}_{\mathrm{IN11}})^{-1} \mathbf{Z}^*_{\mathrm{IN21}} \,. \tag{56}$$

If the lossless linear embedding is invertible, $\mathbf{Z}_{\mathrm{IN21}}$ is square and invertible, and we can use (44)–(45) and (55) in (56) to obtain $\mathbf{Z}_{\mathrm{L}} = \mathbf{Z}^*_{\mathrm{A}}$. Thus, under our assumptions, the available power at the ports of the NMAA during reception, is equal to the available power at the ports of the OMAA during reception.

Accordingly, for an invertible lossless linear embedding, we can assert that:

- in a given direction and for a given polarization, the partial absolute effective areas of the NMAA and OMAA are equal; and
- in a given direction, the absolute effective areas of the NMAA and OMAA are equal, the mean absolute equivalent areas of the NMAA and OMAA are equal, and the minimum absolute equivalent areas of the NMAA and OMAA are equal.

We have established an invariance of the maximum partial absolute gain, mean partial absolute gain, maximum absolute gain, mean absolute gain, minimum absolute gain, partial absolute effective area, absolute effective area, mean absolute equivalent area and minimum absolute equivalent area under an invertible lossless linear embedding.

This invariance under invertible lossless linear embedding clarifies the effect of an invertible lossless linear embedding of the MAA ports. While leaving the invariant parameters unchanged, it may be used to modify other properties of the MAA, such as the impedance matrix, the partial absolute gain patterns for particular excitations, the absolute gain patterns for particular excitations, and the patterns of the parameters of Table 1 to Table 3 that include “reached” in their names.

We have not assumed that the inserted network is reciprocal. If the inserted network is reciprocal, $\mathbf{Z}_{\mathrm{IN}}$ is symmetric and $\mathbf{Z}_{\mathrm{IN}}$ is imaginary because $H(\mathbf{Z}_{\mathrm{IN}}) = 0$. An invertible lossless linear embedding implementing a reciprocal inserted network can be designed to provide decoupling and matching, that is to say a wanted diagonal $\mathbf{Z}_{\mathrm{NA}}$, and predefined radiation patterns for each single-port excitation [24, Sec. II].

## VIII. CONCLUSION

We have studied the parameters of Table 1 to Table 3 in the case of a simple 6-port MAA, and highlighted the fact that, in a given direction, $G_{\mathrm{a\,MAX}}$ and $G_{\mathrm{r\,MAX}}$ may be viewed as envelopes of values of $G_{\mathrm{a}}$ and $G_{\mathrm{r}}$, respectively.

Invariance is a fundamental concept in physics. We proved the invariance of the parameters of Table 2 under a change of excitation variable, and the invariance of some of the parameters of Table 2 and Table 3 under an invertible lossless linear embedding. These invariance properties, like the invariance under a change of experimental system considered in [2, Sec. X], are important properties. Together with the relationships (15)–(22) for a reciprocal MAA, they prove the relevance of the definitions of Section III to the characterization of an MAA for emission and reception.

## APPENDIX A

In the direction of a real wave vector $\mathbf{k}$, a wave polarization may be specified using a polarization vector $\mathbf{u}_{\mathrm{pol}}$, which is a dimensionless complex unit vector. The polarization vector is orthogonal to the given direction, so that $\mathbf{k} \cdot \mathbf{u}_{\mathrm{pol}} = 0$. In [2, Sec. IV], the polarization vector was described as being “of the given wave polarization”. Unfortunately, it became apparent that the wording “of the given wave polarization” is vague and ambiguous.

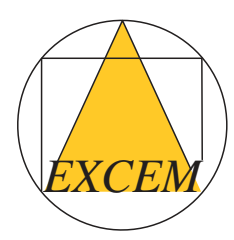

This Appendix A clarifies the definition of the polarization vector $\mathbf{u}_{\rm pol}$ used in [2] and in this article, and takes this opportunity to correct a wrong statement of [2, Sec. III.E] about the polarization mismatch factor.

In a specified direction and for a specified polarization vector $\mathbf{u}_{\rm pol}$ that is orthogonal to the specified direction, we consider a reciprocal single-port remote antenna located at a large enough distance $r_{\rm R}$ of the MAA in the specified direction, the single-port remote antenna being such that its vector effective length in the direction of the MAA, denoted by $\mathbf{h}_{\rm R1}$, is nonzero and satisfies [2, Eq. (34)], that is to say:

$$\mathbf{u}_{\rm pol} = \frac{\mathbf{h}_{\rm R1}}{||\mathbf{h}_{\rm R1}||} . \tag{57}$$

This single-port remote antenna is used to define all "partial" parameters of the MAA in the specified direction:

(a) the partial absolute gain $G_{\rm pa}(\mathbf{u}_{\rm pol})$, its maximum value $G_{\rm pa\,MAX}(\mathbf{u}_{\rm pol})$ and its mean value $G_{\rm pa\,MEA}(\mathbf{u}_{\rm pol})$;
(b) the partial reached gain $G_{\rm pr}(\mathbf{u}_{\rm pol})$, its maximum value $G_{\rm pr\,MAX}(\mathbf{u}_{\rm pol})$ and its mean value $G_{\rm pr\,MEA}(\mathbf{u}_{\rm pol})$;
(c) the partial absolute effective area $A_{\rm pa}(\mathbf{u}_{\rm pol})$; and
(d) the partial reached effective area $A_{\rm pr}(\mathbf{u}_{\rm pol})$.

We observe that, $\mathbf{u}_{\rm pol}$ lying in a plane orthogonal to the specified direction, an orthogonal basis of this plane is such that the coordinates of $\mathbf{u}_{\rm pol}$ in this basis form the normalized complex Jones vector of the electromagnetic field that would be produced in the direction of the MAA, at a large distance of the MAA, by the single-port remote antenna operating in free space [25, Sec. 8.13.2].

In the case of the partial parameters for emission listed in (a) and (b), the single-port remote antenna is used to measure an electric field $\mathbf{E}_{\rm AR}$ produced by the MAA; $G_{\rm pa}(\mathbf{u}_{\rm pol})$ is determined by

$$G_{\rm pa}(\mathbf{u}_{\rm pol}) = \frac{4\pi r_{\rm R}^2}{\eta\, P_{\rm ARP1}}\, e_{\rm pol}^2\, ||\mathbf{E}_{\rm AR}||^2 , \tag{58}$$

where $P_{\rm ARP1}$ is the average power received by the ports of the MAA and

$$e_{\rm pol} = \frac{|\mathbf{h}_{\rm R1} \cdot \mathbf{E}_{\rm AR}|}{||\mathbf{h}_{\rm R1}||\; ||\mathbf{E}_{\rm AR}||} ; \tag{59}$$

and $G_{\rm pr}(\mathbf{u}_{\rm pol})$ is determined by

$$G_{\rm pr}(\mathbf{u}_{\rm pol}) = \frac{4\pi r_{\rm R}^2}{\eta\, P_{\rm AAVG}}\, e_{\rm pol}^2\, ||\mathbf{E}_{\rm AR}||^2 , \tag{60}$$

where $P_{\rm AAVG}$ is the available power of the MGA coupled to the ports of the MAA.

In contrast to what is said in [2, Sec. III.E], the polarization mismatch factor [18], [26, Sec. 5.2], [27, Sec. 16.5] of the single-port remote antenna is $e_{\rm pol}^2$.

In the case of the partial parameters for reception $A_{\rm pa}(\mathbf{u}_{\rm pol})$ and $A_{\rm pr}(\mathbf{u}_{\rm pol})$ in the specified direction, the single-port remote antenna is used to generate an electric field $\mathbf{E}_{\rm B0}$ received by the MAA; by [2, Eq. (31)] and (57), there exists a complex number $\xi_{\rm B0}$, such that

$$\mathbf{E}_{\rm B0} = \xi_{\rm B0}\, \mathbf{u}_{\rm pol} , \tag{61}$$

$\xi_{\rm B0}$ having the dimensions of electric field; $A_{\rm pa}(\mathbf{u}_{\rm pol})$ is given by

$$A_{\rm pa}(\mathbf{u}_{\rm pol}) = \eta\, \frac{P_{\rm BAVA}}{||\mathbf{E}_{\rm B0}||^2} , \tag{62}$$

where $P_{\rm BAVA}$ is the available power of the MAA; and $A_{\rm pr}(\mathbf{u}_{\rm pol})$ is given by

$$A_{\rm pr}(\mathbf{u}_{\rm pol}) = \eta\, \frac{P_{\rm BDP1}}{||\mathbf{E}_{\rm B0}||^2} , \tag{63}$$

where $P_{\rm BDP1}$ is the average power delivered by the ports of the MAA to the MLA.

All partial parameters of the MAA in the specified direction are independent of the choice of the reciprocal single-port remote antenna that satisfies (57). We see that the wording "of the given wave polarization" in [2] actually means that we use a reciprocal single-port remote antenna satisfying (57) to define all partial parameters of the MAA for the specified $\mathbf{u}_{\rm pol}$ and the specified direction.

We also note that, for any $\alpha \in \mathbb{R}$, replacing the polarization vector $\mathbf{u}_{\rm pol}$ with a polarization vector $\mathbf{u}'_{\rm pol} = e^{j\alpha}\mathbf{u}_{\rm pol}$ does not change any of the partial parameters of the MAA in the specified direction.

Now, it follows from the discussion of the Cauchy-Schwarz inequality for positive definite hermitian sesquililear forms, at the end of Appendix A of [2], and from (57)–(60), that each of the partial parameters for emission listed in (a) and (b) is maximized if $\mathbf{u}_{\rm pol}$ is chosen in such a way that there exists a complex number $\xi_{\rm AR}$ having the dimensions of electric field, $\xi_{\rm AR}$ being such that

$$\mathbf{E}_{\rm AR} = \xi_{\rm AR}\, \overline{\mathbf{u}_{\rm pol}} , \tag{64}$$

where the vector $\overline{\mathbf{u}_{\rm pol}}$ is the conjugate of the vector $\mathbf{u}_{\rm pol}$.

In the International Electrotechnical Vocabulary, part 712 relating to single-port antennas, the polarization of an antenna (in a given direction) is defined as "that polarization of the wave radiated by an antenna in the far field region and in a specified direction", and the receiving polarization (of an antenna, in a given direction) is defined as "that polarization of a plane wave of given power flux density incident from a specified direction, which results in maximum received power at the antenna terminals, for that direction" [3]–[4]. Similar definitions are included in [18]. Accordingly, (57) and (61) tell us that $\mathbf{u}_{\rm pol}$ indicates the polarization of the single-port remote antenna. In contrast, the condition (64) for the maximization of the partial parameters for emission listed in (a) and (b) shows that $\mathbf{u}_{\rm pol}$ need not correspond to the receiving polarization of the single-port remote antenna. This fact might at first glance look unexpected from a polarization vector $\mathbf{u}_{\rm pol}$ defined as being "of the given wave polarization". This is why, at the beginning of this Appendix A, the wording "of the given wave polarization" was said to be vague and ambiguous.

However, like (61), (64) is a direct consequence of the fact that we use a reciprocal single-port remote antenna satisfying (57) to define all partial parameters of the MAA for the specified $\mathbf{u}_{\rm pol}$ and the specified direction.

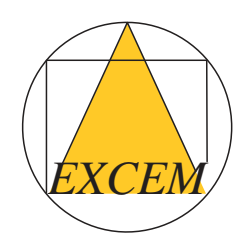

## APPENDIX B

This Appendix B provides a short explanation on the simulation results presented in Section V about an array of six center-fed cylindrical dipole antennas. They are excited by a delta-gap source when they are used for emission.

The simulation are obtained using a method-of-moment-based program implementing the computation technique presented in [26, Ch. 2]. According to this approach, we solve Hallén integral equation, and the numerical solutions are obtained using Lagrange polynomials of order 2 for the basis functions, and point-matching.

Between points belonging to the same dipole antenna, we implement an accurate approximation of the exact thin-wire kernel proposed in [27, Sec. 24.7]. A part of the kernel, comprising all ill-behaved terms, is integrated exactly using known primitives, the other terms of the kernel are integrated numerically. Between points belonging to different dipole antennas, we implement the same approximate thin-wire kernel as the one used in [26, Ch. 2].

We use 32 basis functions for each dipole antenna. Using the symmetry with respect to the plane $z = 0$, we only have to invert a matrix of size 96 by 96 to evaluate the current distributions in each MAA of Section V. They are used to obtain an impedance matrix of size 6 by 6, and $\mathbf{h}_\mathrm{A} = \mathbf{h}_\mathrm{B}$ of size 3 by 6 as a function of the angles $\theta$ and $\varphi$, for each of these MAAs. The impedance matrix is very close to a symmetric matrix. To enforce an exact reciprocity, the symmetric part of this matrix is assumed to be equal to $\mathbf{Z}_\mathrm{A}$.

## APPENDIX C

This Appendix C is about some properties of MAAs made of $N$ thin vertical cylindrical antennas, such as the MAAs considered in Section V and shown in Fig. 1.

Let $\mathcal{B}_\mathrm{S} = (\mathbf{u}_r, \mathbf{u}_\theta, \mathbf{u}_\varphi)$ be the local orthonormal basis of the spherical coordinate system $(r, \theta, \varphi)$. For the MAAs considered here, the currents flowing anywhere in the antennas are vertical. It follows that, in every direction, using the basis $\mathcal{B}_\mathrm{S}$, only the second row of $\mathbf{h}_\mathrm{A}$ contains nonzero entries, so that we can for instance write

$$\mathbf{h}_\mathrm{A} = \begin{pmatrix} 0 & 0 & 0 & 0 & 0 & 0 \\ h_{\mathrm{A}21} & h_{\mathrm{A}22} & h_{\mathrm{A}23} & h_{\mathrm{A}24} & h_{\mathrm{A}25} & h_{\mathrm{A}26} \\ 0 & 0 & 0 & 0 & 0 & 0 \end{pmatrix} \quad (65)$$

in the case $N = 6$ of Section V.

We have $\operatorname{rank} \mathbf{h}_\mathrm{A} \leqslant 1$, and, by (1), $\operatorname{rank} \mathbf{N}_\mathrm{A} \leqslant 1$. Let $\mathbf{h}_\mathrm{AV}$ be the column vector such that $\mathbf{h}_\mathrm{AV}^\mathrm{T}$ is the second row of $\mathbf{h}_\mathrm{A}$, that is

$$\mathbf{h}_\mathrm{A} = \begin{pmatrix} \mathbf{0} \\ \mathbf{h}_\mathrm{AV}^\mathrm{T} \\ \mathbf{0} \end{pmatrix}. \quad (66)$$

It follows from (1) that

$$\mathbf{N}_\mathrm{A} = \frac{\pi\eta}{\lambda^2}\, \overline{\mathbf{h}_\mathrm{AV}} \mathbf{h}_\mathrm{AV}^\mathrm{T}\,, \quad (67)$$

where $\overline{\mathbf{h}_\mathrm{AV}}$ is the conjugate of $\mathbf{h}_\mathrm{AV}$. Since

$$\underset{\sim}{\mathbf{u}}_\mathrm{pol}^\mathrm{T} \mathbf{h}_\mathrm{A} = (\underset{\sim}{\mathbf{u}}_\mathrm{pol})_2\, \mathbf{h}_\mathrm{AV}^\mathrm{T} = (\mathbf{u}_\mathrm{pol} \cdot \mathbf{u}_\theta)\, \mathbf{h}_\mathrm{AV}^\mathrm{T}\,, \quad (68)$$

it follows from (5) that

$$\mathbf{N}_\mathrm{Ap}(\mathbf{u}_\mathrm{pol}) = \frac{\pi\eta}{\lambda^2}\, |\mathbf{u}_\mathrm{pol} \cdot \mathbf{u}_\theta|^2\, \overline{\mathbf{h}_\mathrm{AV}} \mathbf{h}_\mathrm{AV}^\mathrm{T} \quad (69)$$

and from (67) that

$$\mathbf{N}_\mathrm{Ap}(\mathbf{u}_\mathrm{pol}) = |\mathbf{u}_\mathrm{pol} \cdot \mathbf{u}_\theta|^2\, \mathbf{N}_\mathrm{A}\,. \quad (70)$$

Using $\operatorname{rank} \mathbf{N}_\mathrm{A} \leqslant 1$ and (70), we obtain:

$$G_\mathrm{a\,MEA} = \frac{G_\mathrm{a\,MAX}}{N}\,, \quad (71)$$

$$(N > 1) \Longrightarrow (G_\mathrm{a\,MIN} = 0)\,, \quad (72)$$

$$G_\mathrm{pa\,MAX}(\mathbf{u}_\mathrm{pol}) = |\mathbf{u}_\mathrm{pol} \cdot \mathbf{u}_\theta|^2\, G_\mathrm{a\,MAX}\,, \quad (73)$$

$$G_\mathrm{pa\,MEA}(\mathbf{u}_\mathrm{pol}) = |\mathbf{u}_\mathrm{pol} \cdot \mathbf{u}_\theta|^2\, G_\mathrm{a\,MEA}\,, \quad (74)$$

$$G_\mathrm{r\,MEA} = \frac{G_\mathrm{r\,MAX}}{N}\,, \quad (75)$$

$$(N > 1) \Longrightarrow (G_\mathrm{r\,MIN} = 0)\,, \quad (76)$$

$$G_\mathrm{pr\,MAX}(\mathbf{u}_\mathrm{pol}) = |\mathbf{u}_\mathrm{pol} \cdot \mathbf{u}_\theta|^2\, G_\mathrm{r\,MAX} \quad (77)$$

and

$$G_\mathrm{pr\,MEA}(\mathbf{u}_\mathrm{pol}) = |\mathbf{u}_\mathrm{pol} \cdot \mathbf{u}_\theta|^2\, G_\mathrm{r\,MEA}\,. \quad (78)$$

The MAAs considered here being reciprocal, we have $\mathbf{h}_\mathrm{B} = \mathbf{h}_\mathrm{A}$, so that

$$\mathbf{h}_\mathrm{B}^\mathrm{T}\, \underset{\sim}{\mathbf{u}}_\mathrm{pol} = (\mathbf{u}_\mathrm{pol} \cdot \mathbf{u}_\theta)\, \mathbf{h}_\mathrm{AV} \quad (79)$$

and

$$\mathbf{h}_\mathrm{B}^\mathrm{T}\, \mathbf{P} = \begin{pmatrix} \mathbf{h}_\mathrm{AV} & \mathbf{0} \end{pmatrix}. \quad (80)$$

Using (79) in (10) and (13), we get

$$A_\mathrm{pa}(\mathbf{u}_\mathrm{pol}) = \frac{\eta}{4}\, |\mathbf{u}_\mathrm{pol} \cdot \mathbf{u}_\theta|^2\, \mathbf{h}_\mathrm{AV}^*\, H(\mathbf{Z}_\mathrm{A})^{-1}\, \mathbf{h}_\mathrm{AV} \quad (81)$$

and

$$A_\mathrm{pr}(\mathbf{u}_\mathrm{pol}) = \eta\, |\mathbf{u}_\mathrm{pol} \cdot \mathbf{u}_\theta|^2\, \mathbf{h}_\mathrm{AV}^* \mathbf{Y}_\mathrm{AL}^* H(\mathbf{Z}_\mathrm{L}) \mathbf{Y}_\mathrm{AL} \mathbf{h}_\mathrm{AV}\,. \quad (82)$$

Using (80) in (12) and (14), we get the 2 by 2 matrices

$$\mathbf{N}_\mathrm{Ba} = \frac{\eta}{4} \begin{pmatrix} \mathbf{h}_\mathrm{AV}^* H(\mathbf{Z}_\mathrm{A})^{-1} \mathbf{h}_\mathrm{AV} & 0 \\ 0 & 0 \end{pmatrix} \quad (83)$$

and

$$\mathbf{N}_\mathrm{Br} = \eta \begin{pmatrix} \mathbf{h}_\mathrm{AV}^* \mathbf{Y}_\mathrm{AL}^* H(\mathbf{Z}_\mathrm{L}) \mathbf{Y}_\mathrm{AL} \mathbf{h}_\mathrm{AV} & 0 \\ 0 & 0 \end{pmatrix}, \quad (84)$$

which are diagonal. It follows from (81)–(82), or from (83)–(84), that

$$A_\mathrm{a} = \frac{\eta}{4}\, \mathbf{h}_\mathrm{AV}^* H(\mathbf{Z}_\mathrm{A})^{-1} \mathbf{h}_\mathrm{AV}\,, \quad (85)$$

$$A_\mathrm{aeq\,MIN} = 0\,\mathrm{m}^2\,, \quad (86)$$

$$A_\mathrm{aeq\,MEA} = \frac{1}{2}\, A_\mathrm{a}\,, \quad (87)$$

$$A_\mathrm{r} = \eta\, \mathbf{h}_\mathrm{AV}^* \mathbf{Y}_\mathrm{AL}^* H(\mathbf{Z}_\mathrm{L}) \mathbf{Y}_\mathrm{AL} \mathbf{h}_\mathrm{AV}\,, \quad (88)$$

$$A_\mathrm{req\,MIN} = 0\,\mathrm{m}^2 \quad (89)$$

and

$$A_\mathrm{req\,MEA} = \frac{1}{2}\, A_\mathrm{r}\,. \quad (90)$$

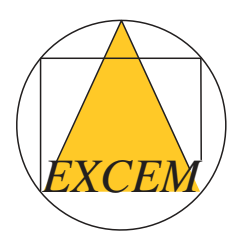

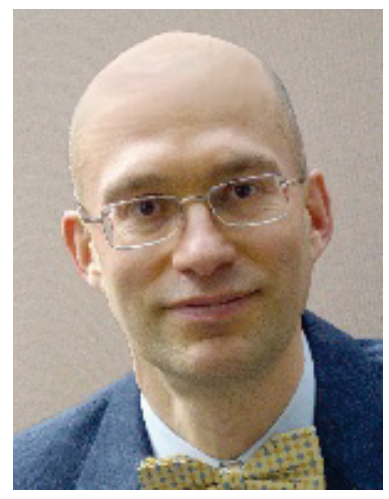

FRÉDÉRIC BROYDÉ was born in France in 1960. He received the M.S. degree in physics engineering from the Ecole Nationale Supérieure d'Ingénieurs Electriciens de Grenoble (ENSIEG) and the Ph.D. in microwaves and microtechnologies from the Université des Sciences et Technologies de Lille (USTL).

He co-founded the Excem corporation in May 1988, a company providing engineering and research and development services. He is president of Excem since 1988. He is now also president and CTO of Eurexcem, a subsidiary of Excem. Most of his activity is allocated to engineering and research in electronics, radio, antennas, electromagnetic compatibility (EMC) and signal integrity.

Dr. Broydé is author or co-author of about 100 technical papers, and inventor or co-inventor of about 90 patent families, for which 71 patents of France and 49 patents of the USA have been granted. He became a Senior Member of the IEEE in 2001 and is now a Life Senior Member of the IEEE. He is a licensed radio amateur (F5OYE).

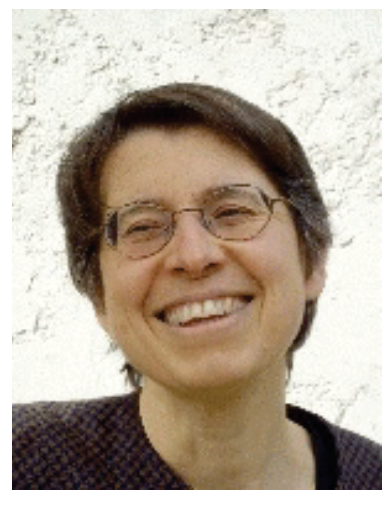

EVELYNE CLAVELIER was born in France in 1961. She received the M.S. degree in physics engineering from the Ecole Nationale Supérieure d'Ingénieurs Electriciens de Grenoble (ENSIEG).

She is co-founder of the Excem corporation, based in Maule, France, and she is currently CEO of Excem. She is also president of Tekcem, a company selling or licensing intellectual property rights to foster research. She is an active engineer and researcher.

Her current research areas are radio communications, antennas, matching networks, EMC and circuit theory.

Prior to starting Excem in 1988, she worked for Schneider Electrics (in Grenoble, France), STMicroelectronics (in Grenoble, France), and Signetics (in Mountain View, CA, USA).

Ms. Clavelier is the author or a co-author of about 90 technical papers. She is co-inventor of about 90 patent families. She is a Senior Member of the IEEE since 2002. She is a licensed radio amateur (F1PHQ).

•••